\documentclass{aa}  
\usepackage{graphicx}
\usepackage{txfonts}
\usepackage{longtable}
\usepackage{longtable,lscape}
\usepackage[figuresright]{rotating}
\newcommand{\msun}{$\,M_\odot$}

\newcommand{\kms}{\ensuremath{\,{\rm km\,s}^{-1}}}
\newcommand{\rgal}{\ensuremath{{R_{\rm Gal}}}}
\newcommand{\dhel}{\ensuremath{{d_{\rm hel}}}}

\newcommand{\ciii}{{\rm C\,}{{\sc iii}}}
\newcommand{\oiii}{{\rm O\,}{{\sc iii}}}

\newcommand{\nii}{{\rm N\,}{{\sc ii}}}

\begin{document}
\title{Bayesian posterior classification of planetary nebulae according to the Peimbert types
\thanks{Table~\ref{tab:data} and Table~\ref{tab:probab} are only available in electronic at the CDS via 
anonymous ftp to cdsarc.u-strasbg.fr (130.79.125.5) 
or via http://cdsweb.u-strasbg.fr/Abstract.html}
}

\author{C. Quireza\inst{1}, H.J. Rocha-Pinto\inst{2} and W.J. Maciel\inst{3}}

\offprints{C. Quireza}

\institute{Observat\'orio Nacional, Rua General Jos\'e Cristino 77, 20921-400, Rio de Janeiro, RJ, Brazil\\ 
\email{quireza@on.br}
\and Observat\'orio do Valongo, Universidade Federal do Rio de Janeiro, Lad. Pedro Ant\^onio 43, 20080-090 Rio de Janeiro RJ, Brazil\\
\email{helio@ov.ufrj.br}
\and Instituto de Astronomia, Geof\'{\i}sica e Ci\^encias Atmosf\'ericas, Universidade de S\~ao Paulo, Rua do Mat\~ao 1226, 05508-900, S\~ao Paulo, SP, Brazil\\
\email{maciel@astro.iag.usp.br}}

\date{Received 14 June 2007 / Accepted 30 August 2007}

% \abstract{}{}{}{}{}, 5 {} token are mandatory
 
\abstract
  % context heading (optional), leave it empty if necessary  
{Galactic planetary nebulae are observed with a wide variety of kinematic properties, spatial distribution, chemical composition and  morphologies, comprising members of the dominant stellar populations of our Galaxy. Due to their broad astrophysical interest, a proper characterization of these populations is of major importance.}
{In this paper we present a re-analysis of the criteria used to characterize the Peimbert classes I, IIa, IIb, III and IV, through a statistical study of a large sample of planetary nebulae previously classified according to these groups. In the original classification, it is usual to find planetary nebulae that cannot be associated with a single type; these most likely have dubious classifications into two or three types. Statistical methods can greatly contribute in providing a better characterization of planetary nebulae groups.}
{We use the Bayes Theorem to calculate the posterior probabilities for an object to be member of each of the types I, IIa, IIb, III and IV. This calculation is particularly important for planetary nebulae that are ambiguously classified in the traditional method. The posterior probabilities are defined from the probability density function of classificatory parameters of a well-defined sample, composed only by planetary nebulae unambiguously fitted into the Peimbert types. Because the probabilities depend on the available observational data, they are conditional probabilities, and, as new observational data are added to the sample, the classification of the nebula can be improved, to take into account this new information.}
{This method differs from the original classificatory scheme, because it provides a quantitative result of the representativity of the object within its group. Also, through the use of marginal distributions it is possible to extend the Peimbert classification even to those objects for which only a few classificatory parameters are known.}
{We found that ambiguities in the classification of planetary nebulae into the Peimbert types, should be associated to difficulties in defining sharp boundaries for the progenitor star mass for each of these types. Those can be at least partially explained by real overlaps of some of the parameters that characterize the different stellar populations. Those results suggest the need of a larger number of classificatory parameters for a reliable physical classification of planetary nebulae.}


\titlerunning{A reanalysis of the Peimbert planetary nebulae types}
\authorrunning{C. Quireza et al.}
\maketitle

\section{Introduction}\label{sec:intro}

Planetary nebulae (PNe) result from the ejection of the external envelope of low and intermediate mass stars during their evolution from the asymptotic giant branch (AGB) to the white dwarf stage. Roughly speaking, the nebula will be created by the interaction of the slow AGB wind with the new fast stellar wind from a post-AGB phase, which will sweep up the circumstellar envelopes into a shell-like structure (Kwok et al. 1978; Kwok 1982, 1983; Kahn 1983 and Volk \& Kwok 1985). The energy source of the nebula is the ultraviolet radiation from the hot central star, which gradually ionizes the shell, at the same time the shell grows in mass as more AGB wind material is swept up. 

Because PNe come from stars in a large range of main sequence masses (0.8--8.0~\msun, cf. Peimbert 1990), they are observed with a wide variety of spatial and kinematic properties, chemical composition, and morphology, comprising a mixed group of objects ranging from Population I (slow moving disk objects) to extreme Population II (fast moving objects from the bulge and the halo). On account of this, PNe are objects of large astrophysical interest. Their application is multidisciplinary: they probe the evolution of low and intermediate mass stars, and the Galactic nucleosynthesis history. Elements that are not modified by the nucleosynthesis of the parent star allow us to investigate the composition of the interstellar medium at the time when the star formed. PNe are useful tracers of Galactic kinematics (Schneider \& Terzian 1983; Durand et al. 1998; Maciel \& Lago 2005), Galactic chemical evolution (cf. Costa \& Maciel 2006), and are even useful tools in the investigation of some cosmological problems (see Balser et al. 1997, 1999, 2006, and references therein about the measurement of $^3$He abundance in PNe and astrophysical implications). This richness of information generally requires a proper selection of the objects to be used in a specific analysis. By selecting PNe that closely reflect the properties of the interstellar medium out of which their central stars have been formed, we can study, for instance, the presence of abundance gradients in the disk of the Galaxy. There is not an ideal method of performing such selection, since how rigorous one is depends on the kind of study to be performed and of the goal of the analysis. However, it is important to understand the different groups (or populations) of PNe and, at least, some of their main characteristics.

Among the different planetary nebula classification schemes found in the literature (Greig 1971, 1972; Kaler 1983; Heap \& Augensen 1987), one of the most efficient is probably that proposed by Peimbert (1978), which suggested the classification of PNe into four types, namely I, II, III and IV (see also Peimbert \& Serrano 1980; Peimbert \& Torres-Peimbert 1983). Each of these groups should roughly correspond to a stellar population of the Galaxy. These groups also represent a sequence of mass intervals for the planetary nebula progenitor star, type I representing the youngest population, with the most massive progenitors, and type IV the oldest population, with the less massive progenitor stars. Later a type V was also defined to include bulge objects (Maciel 1989), which should constitute a distinct stellar population, but the selection criteria are different than those used to select the four original types. In order to characterize the types I to IV, Peimbert took into account, chemical, kinematical and spatial properties of PNe. Such criteria were supplemented by Peimbert \& Torres-Peimbert (1983), Fa\'undez-Abans \& Maciel (1987a) and Maciel (1989), and have been continuously used in Galactic evolution related issues, mainly in the study of the abundance gradients (Maciel \& Fa\'undez-Abans 1985; Fa\'undez-Abans and Maciel 1987b; Maciel \& Quireza 1999; Maciel 2002; Maciel et al. 2003, 2006; Perinotto \& Morbidelli 2006). The classification, however, has its flaws: some criteria are poorly defined, and there is a tendency to find objects with different characteristics in the same class; moreover, several objects simply do not fit in any of the existing classes, which suggests the need for a larger number of classificatory parameters, and/or more rigorous classification criteria.

In this paper we present a re-analysis of the criteria used to characterize the Peimbert classes, through a statistical study of a large sample of PNe previously classified into the types I to IV. The process of selection of the fundamental parameters is described in \S\ref{sec:data}. In \S\ref{sec:preclass} we describe the classification criteria by Peimbert (1978) and we present the classification of our PNe according to such criteria. In \S\ref{sec:posclass} we present our statistical analysis, which provides us information about how well each of the objects accommodates in a given class. We discuss the results and implications of our analysis in \S\ref{sec:discuss}. In \S\ref{sec:concl} we summarize our main results.

\section{Planetary nebulae data base} \label{sec:data}

Ideally, when analyzing a number of objects, it is desirable to use a homogeneous sample, with data obtained using similar techniques and self-consistent measurements, which results in very accurate information. Unfortunately, samples with precise and self-consistent values of abundance, distances, and other fundamental parameters of PNe are still not large enough for statistical studies. We have concentrated efforts in generating a large sample of PNe of different properties, so we can characterize the various populations/groups of PNe, belonging to the different structures of our Galaxy. In order to do that, we have compiled the nebular properties from selected sources from the literature. 

Our sample consists of PNe for which abundances have been determined in the literature. Most of it consists of objects from the Strasbourg-ESO Catalogue of Galactic Planetary Nebulae (Acker et al. 1992), plus new observed PNe, not present in the Acker et al. catalogue (Kohoutek 1994; Beaulieu et al. 1999; Escudero \& Costa 2001). The resulting sample has a total of 476 objects. The fundamental parameters we looked for were the helium, nitrogen and oxygen abundances relative to hydrogen, the heliocentric distance, \dhel, and the radial velocity relative to the Local Standard of Rest, $V_{\rm LSR}$. Additionally, we have compiled the flux at 5~GHz, $S_{\rm 5~GHz}$, and the angular diameter of the PNe, $\Theta_{\rm diam}$, which have been used to exclude bulge objects, as will be discussed in \S~\ref{sec:preclass}. The adopted parameters are listed in Table~\ref{tab:data}, which is available only in electronic form. The process of selection of these parameters is described in the subsections ahead. In the table, for each planetary nebula, we also list the common name, the PNG number (Acker et al. 1992), and calculated properties, namely the Galactocentric distance, \rgal, the height in relation to the Galactic plane, $z$, and the peculiar velocity, $\Delta V$. Explicit formulae for these quantities are given later on in \S\ref{sec:peimbert}. The resulting Peimbert classes are listed in the last column. Those quantities will be commented in \S~\ref{sec:preclass}. For units, the distances are given in kpc, the angular size in arcsec, and the  flux in mJy. The oxygen, and nitrogen abundances are given as $\log(\mathrm{X/H}) + 12$, as usual.  The corresponding references for each of these quantities are provided in the table. Most of the Galactic coordinate values come from the Acker et al. (1992) catalogue.

\subsection{Elemental abundance} \label{sec:abund}

For each planetary nebula, we have tried to compile all nebular abundance measurements available in the literature. The sample should be as complete as we could make it up to 2004. Additional abundances published after 2004 have been occasionally included, such as the ISO (Infrared Space Observatory) results from Pottasch \& Bernard-Salas (2006). For each object, we have kept only those abundance determinations that agree within the adopted uncertainties, which we estimate to be in the range 0.1-0.4~dex for most objects. The adopted abundances are the average of all the remaining abundances for a given PNe. Abundances from Costa et al. (1996, 1997, 2004), Cuisinier et al. (2000), Escudero \& Costa (2001), and Escudero et al. (2004) were preferred to be used individually or averaged. We did so because such studies used the same instruments, similar observational techniques and data analysis procedures. Pottasch \& Bernard-Salas (2006) were also included in this last group, since their abundance measurements should be one of the most accurate available in the literature up to now.

%
%
%=====================================================================
% Table 1 available electronically only
\begin{landscape}
\begin{table}
\caption{\label{tab:data} Parameters and pre-classification (full table and reference lists are available electronically)}
\begin{tabular}{llrlrlrlrrlrrrrrlrll}
\hline\hline
\noalign{\smallskip}
Name & PNG & He/H & Ref. & $\epsilon({\rm O})^{\mathrm{\dagger}}$ & Ref. & $\epsilon({\rm N})^{\mathrm{\dagger}}$ & Ref. &  $\log({\rm N/O})$ & \dhel & Ref. & \rgal & $\vert z \vert$ & $V_{\rm LSR}^{\mathrm{\ddagger}}$ & $\vert \Delta V \vert$ & $\Theta_{\rm diam}$ & Ref. & $S_{\rm 5~GHz}$ & Ref. &  Type \\
     &     &      &      &     &      &            &     &       & (kpc) &  & (kpc) & (kpc) & (km/s)                                                                                                                                                                                                                                                                                                 & (km/s)     & (arcsec) &      & (mJy) &     &       \\
\noalign{\smallskip}
\hline
\noalign{\smallskip}
A4             &    144.3$-$15.5      &  \dots    &                 \dots    &  \dots    &                    \dots    &  \dots    &                 \dots    &  $-$0.17    &  6.1    &    10    &  12.8    &   1.630    &     \dots    &   \dots    &    20.0    &        27    &     1.5    &       9    &                  indef.      \\
A12            &    198.6$-$06.3      &  0.119    &                     3    &   8.93    &                        3    &   8.06    &                     3    &  $-$0.87    &  2.0    &     4    &   9.5    &   0.219    &     \dots    &   \dots    &    37.0    &        18    &    36.0    &       9    &                    IIbD      \\
A18            &    216.0$-$00.2      &  0.152    &                     3    &   7.99    &                        3    &   7.97    &                     3    &  $-$0.02    &  1.6    &     4    &   8.9    &   0.006    &     \dots    &   \dots    &    73.0    &         7    &    17.0    &       9    &                      ID      \\
A20            &    214.9$+$07.8      &  0.125    &                     3    &   8.80    &                        3    &  \dots    &                 \dots    &  \dots      &  2.0    &     4    &   9.3    &   0.271    &     \dots    &   \dots    &    67.0    &         7    &     7.0    &       9    &               (I/IIa)D*      \\
A24            &    217.1$+$14.7      &  \dots    &                 \dots    &  \dots    &                    \dots    &  \dots    &                 \dots    &  $+$0.43    &  0.3    &     7    &   7.8    &   0.076    &    $+$0.9    &    14.0    &   354.8    &        27    &    36.0    &       9    &                  indef.      \\
A35            &    303.6$+$40.0      &  \dots    &                 \dots    &  \dots    &                    \dots    &  \dots    &                 \dots    &  $-$0.10    &  0.2    &    17    &   7.5    &   0.129    &    $-$5.9    &    15.1    &   772.0    &        27    &   255.0    &       9    &                  indef.      \\
A50            &    078.5$+$18.7      &  0.089    &                    21    &  \dots    &                    \dots    &  \dots    &                 \dots    &  $-$0.40    &  2.8    &    10    &   7.5    &   0.898    &  $-$145.2    &   175.2    &    27.0    &        27    &     1.0    &       9    &                    IIaC      \\
A65            &    017.3$-$21.9      &  0.260    &                    22    &   8.18    &                       22    &   7.33    &                    22    &  $-$0.85    &  1.5    &    10    &   6.3    &   0.559    &   $+$21.8    &     1.9    &   104.0    &        27    &     4.0    &       9    &                    IIaF      \\
A70            &    038.1$-$25.4      &  0.180    &                    22    &   7.98    &                       22    &   7.52    &                    22    &  $-$0.46    &  3.5    &    10    &   5.5    &   1.501    &   $-$69.0    &   145.0    &    42.0    &        27    &    12.0    &       9    &                    IIaE      \\
A71            &    084.9$+$04.4      &  \dots    &                 \dots    &  \dots    &                    \dots    &  \dots    &                 \dots    &  $+$0.38    &  0.9    &    10    &   7.6    &   0.069    &     \dots    &   \dots    &   158.0    &        27    &    82.8    &       9    &                  indef.      \\
A77            &    097.5$+$03.1      &  \dots    &                 \dots    &   8.41    &                       21    &   7.02    &                    21    &  $-$1.39    &  1.5    &    10    &   7.9    &   0.081    &  $-$103.4    &   122.5    &    65.8    &        27    &   307.6    &       9    &                  indef.      \\
A82            &    114.0$-$04.6      &  \dots    &                 \dots    &  \dots    &                    \dots    &  \dots    &                 \dots    &  $-$0.28    &  2.0    &    10    &   8.6    &   0.160    &   $-$24.6    &    25.4    &    81.0    &        27    &     5.3    &       9    &                  indef.      \\
.              &                      &           &                          &           &                             &           &                          &             &         &          &          &            &              &            &            &              &            &            &                              \\
.              &                      &           &                          &           &                             &           &                          &             &         &          &          &            &              &            &            &              &            &            &                              \\
.              &                      &           &                          &           &                             &           &                          &             &         &          &          &            &              &            &            &              &            &            &                              \\
\noalign{\smallskip}
\hline
\end{tabular}

\begin{list}{}{}
\item[$^{\dagger}$] $\epsilon({\rm X}) = \log ({\rm X/H}) + 12$.
\item[$^{\ddagger}$] Radial velocities from the catalog of Durand et al. (1998). For every object, the heliocentric radial velocities have been converted to the Local Standard of Rest radial velocities (see text).
\end{list}{}{}

\noindent References of abundances.
(3) Costa et al. 2004;
(21) Perinotto 1991;
(22) Perinotto et al. 1994; \dots

\noindent References of distances.
(4) Costa et al. 2004 (see references in);
(7) Harris et al. 1997;
(10) Maciel 1984;
(17) Pottasch 1996; \dots

\noindent References of the angular diameters.
(7) Cahn \& Kaler 1971;
(18) Perek \& Kohoutek 1967;
(27) Zhang 1995 (see references in); \dots

\noindent References of the 5 GHz flux densities.
(9) Zhang 1995 (see references in); \dots

\end{table}
\end{landscape}
%%=====================================================================
%
%

\subsection{Heliocentric distances} \label{sec:dist}

The range in distances as determined by different authors can be very large, and choosing a good distance estimate is a difficult task. Only a small number of PNe have accurate individual estimates of their distances. For the vast majority of PNe, the only distances available are those obtained from statistical methods. Here, whenever possible, we use individual distances over statistical estimates. Those include trigonometric parallax (Harris et al. 1997), Hipparcos parallaxes (Pottasch 1997; Pottasch \& Acker 1998), spectroscopic parallax, radial expansion distances measured by VLA or by Hubble Space Telescope (Hajian et al. 1995; Terzian 1997) and local extinction (Martin 1994). We have chosen as far as possible to use distances estimated from parallax or expansion method, because in both cases the distances are independent of any (assumed) property of the nebula or its central star. In Table~\ref{tab:data}, individual distances represent about 17\% of the sample. Among statistical distances the catalog of Maciel (1984) represent almost half of the distance sample, and is complemented by Cahn et al. (1992) and Zhang (1995). Cahn et al.'s distances also substituted lower and upper limits from Maciel's catalog. For the bulge and anticenter PNe compiled respectively from Escudero et al. (2004) and Costa et al. (2004), original references for the distances are Schneider \& Buckley (1996), van de Steene \& Zijlstra (1995), and Amnuel et al. (1984). We could not find distances for 71 objects in our sample.

\subsection{Radial velocity relative to the Local Standard of Rest} \label{sec:vlsr}

We have used heliocentric radial velocities from the catalog of Durand et al. (1998), which results from the selection of all known Galactic PNe radial velocities up to 1998. For a given object in their catalog, radial velocities are weighted averages of each existing velocity by the inverse of the square of its associated error. According to the authors, 90\% of the total sample have accuracies better than 20~\kms.

The heliocentric radial velocities, $V_{hel}$, have been converted to the Local Standard of Rest radial velocities using the basic solar motion $(u,v,w)=(-9,11,6)$~\kms, as given by Mihalas \& Binney (1981). For a right hand coordinate system the correction is
\begin{equation}
V_{\rm LSR} = V_{hel} + (9 \cos b \cos l + 11 \cos b \sin l + 6 \sin b)~\kms \,, \label{vlsr}
\end{equation}
\noindent were $l$ and $b$ are the Galactic coordinates of the objects. We have radial velocities for 428 PNe in our sample.

\subsection{Angular diameter} \label{sec:size}

Whenever possible, we have used radio continuum measurements over optical ones. We also have tried to give priority to measurements made by a same group of researchers, which generally use the same observational techniques (for instance, Aaquist \& Kwok 1990, 1991; Kwok \& Aaquist 1993). Those choices have been taken as an attempt to make the sample less heterogeneous. In Table~\ref{tab:data} we list angular sizes for 448 objects. For elongated nebulae, a geometric mean of the semi-major and semi-minor axes was used whenever necessary.

\subsection{Radio continuum flux density at 5~GHz} \label{sec:flux}

Most of the radio fluxes have been adopted from Zhang (1995), which used data obtained with the Very Large Array (VLA) (Aaquist \& Kwok 1990; Zijlstra et al. 1989; Gathier et al. 1983), plus some single-dish measurements (Milne \& Aller 1975; Milne 1979) and measurements tabulated by Pottasch (1984) (see Zhang \& Kwok 1993). We list fluxes for 388 objects.

Angular sizes and flux densities have been used with the sole purpose of selecting bulge objects in our sample (\S~\ref{sec:preclass}). These quantities have not been used in our statistical analysis.

\section{Pre-classification of planetary nebulae} \label{sec:preclass}

\subsection{Peimbert types and classification} \label{sec:peimbert}

Peimbert (1978) has divided the Galactic PNe into four types on the basis of their chemical, spatial and kinematic properties. Ideally, each of these groups should correspond to distinct intervals of mass of the progenitor star, whose evolution should affect differently the chemical composition of the ejected planetary nebula envelope (Calvet \& Peimbert 1983; Peimbert \& Serrano 1980; Peimbert \& Torres-Peimbert 1983).

Type I PNe are He and N rich, presenting extremely filamentary structure. They have been defined by Peimbert (1978) as PNe with ${\rm He/H} \geq 0.14$ or $\log{\rm (N/O)} \geq 0$, but later on Peimbert \& Torres-Peimbert (1983) have relaxed this definition by including those objects with ${\rm He/H} \geq 0.125$ or $\log{\rm (N/O)} \geq -0.30$. This excess in He and N can be eventually measured as a N/O excess in the nebula (Peimbert 1990), and has two possible explanations: i) N contamination in the outer atmosphere of the planetary nebula progenitor star, caused by dredge up processes. Such N would be produced at the expenses of C and O during H burning by the CNO cycle; and ii) The progenitor star would have been formed more recently from a medium richer in He and heavy elements. Both causes suggest that type I PNe result from the most massive progenitor stars in the intermediary mass range, being the youngest of all disk PNe. They are associated with the Galactic thin disk (closer to the Galactic plane), whose scale height is, here, assumed to be about 300~pc (Gilmore \& Reid 1983, Reddy et al. 2006). Because thin disk stars orbit the Galactic center in nearly circular orbits, type I PNe should present low velocity dispersion (Dutra \& Maciel 1990; Maciel \& Dutra 1992).

Type II PNe belong to the intermediate disk population. Still associated with the thin disk, they have approximately circular orbits. Members of this group are generally older than type I PNe, having been formed at a time when the interstellar medium was more metal-poor in heavy elements. Therefore, they may present slight underabundances of elements such as O, S and Ne. Moreover, their low mass prevents strong He or N enrichments. According to Peimbert (1978), most existing PNe are probably of type II, having He/H $< 0.125$ and $\log(\mathrm{N/O})< -0.3$. Later on, Fa\'undez-Abans \& Maciel (1987a), further subdivided the type II PNe into types IIa and IIb, according to their N abundance (types IIb having no N enrichment and types IIa having some N enrichment intermediate between type I and type IIb). In such scheme, type IIa have been formed from progenitor stars near the high mass bracket. Because of that, type IIa PNe may present slightly enriched abundances of O, S, Ne, and Ar, and some definite enrichment in He and N.

Here, as in Maciel \& Quireza (1999), we consider a slightly more rigid criteria to sample PNe in the thin disk: we consider as type I only those objects for which both conditions ${\rm He/H} > 0.125$ and $\log({\rm N/O}) > -0.30$ are satisfied. As a consequence, objects for which only one of these conditions is satisfied have been considered as type IIa (Fa\'undez-Abans \& Maciel 1987a), so that they are in fact type II objects. This procedure should assure a correct sampling of the nebulae associated with the more massive progenitor stars, in the high end of the mass interval that characterizes intermediate mass stars. In this interval, which corresponds to type I planetary nebula, the nebular oxygen abundance may be affected by ON cycling in the progenitor stars (Henry 1990; Maciel 1992; Perinotto \& Morbidelli 2006).

%=====================================================================
\setcounter{table}{1}
\begin{table*}
\begin{minipage}[t][]{\textwidth}
%\begin{minipage}[t][180mm]{\textwidth}
\caption{Classification criteria for Peimbert classes}\label{tab:criteria}
\centering
\renewcommand{\footnoterule}{}
\begin{tabular}{llcccccccl}
\hline\hline
\noalign{\smallskip}
\multicolumn{2}{l}{Type} & He/H$^{\mathrm{a}}$ & $\log({\rm N/O})$$^{\mathrm{a}}$ & $\log({\rm N/H}) + 12$ & $\vert z \vert$$^{\mathrm{b}}$ & $\vert \Delta V \vert$ & $M_{ms}$ & Age & Location \\
    &  &      &                        &                   & (kpc)           & (\kms)          & ($M_\odot$)  & (Gy) &         \\
\noalign{\smallskip}
\hline
\noalign{\smallskip}
I & & $\geq 0.125$ & $\geq -0.30$ & & $\ll 1\,(< 0.3)$ & $< 60$ & 2.4--8.0$^{\mathrm{c}}$ & 0--2$^{\mathrm{d}}$ & thin disk \\
\noalign{\smallskip}
\hline
\noalign{\smallskip}
 & IIa & $\geq 0.125$ & $< -0.30$ & $\geq 8.0$ &   &   &   &   &   \\
II & IIa & $< 0.125$ & $\geq -0.60$ & $\geq 8.0$ & $< 1$ & $< 60$ & 1.2--2.4$^{\mathrm{c}}$ & 4--6$^{\mathrm{d}}$ & thin disk \\
 & IIb & $< 0.125$ & $< -0.60$ & $< 8.0$ &   &   &   &   &   \\
\noalign{\smallskip}
\hline
\noalign{\smallskip}
III & &                      &                      & & $\geq 1\,(\leq 1.45)$ & $\ga 60$ & 1.0--1.2$^{\mathrm{c}}$ & 8--10$^{\mathrm{d}}$ & thick disk \\
\noalign{\smallskip}
\hline
\noalign{\smallskip}
IV & &                      &                      & & $\gg 1\,(> 1.45)$ & $\ga 100$ & 0.8--1.0$^{\mathrm{c}}$ & $> 10$ & halo \\
\noalign{\smallskip}
\hline
\noalign{\smallskip}
V & &                      &                      &                     & $< 1.3$ & large range & large range & large range & bulge \\
\noalign{\smallskip}
\hline
\end{tabular}
\vfill
\end{minipage}
\begin{list}{}{}
\item[$^{\mathrm{a}}$] No abundance criteria were originally defined to the types III, IV and V. Average abundances can be calculated using data from Chiappini \& Maciel (1994) for types III and IV, and data from Escudero \& Costa (2001), Cuisinier et al. (2000) and Escudero et al. (2004) for type V. Mean values of He/H and $\log({\rm N/O})$ are respectively 0.099 and -0.68 for type III, 0.104 and -0.67 for type IV and 0.110 and -0.60 for type V PNe.
\item[$^{\mathrm{b}}$] According to the original criteria, type I nebulae have $\vert z \vert \ll 1$ and type III $\vert z \vert \geq 1$. Here, quantitative limits for $\vert z \vert$ were defined assuming the scale height of 0.3~kpc to the thin disk and 1.45~kpc to the thick disk (Gilmore \& Reid 1983). Criteria for type V are defined using $R_\odot = 7.6$~kpc and assuming that the object should be within $10\degr$ from the Galactic center.
\item[$^{\mathrm{c}}$] Maciel (1992)
\item[$^{\mathrm{d}}$] Maciel \& K\"oppen (1994)
\end{list}
\end{table*}
%=====================================================================

Type III PNe have been defined by Peimbert (1978) as those objects with $\vert \Delta V \vert > 60$~\kms\, which do not belong to the halo population. They are high velocity objects from the thick disk population. Such PNe were ejected from stars, which are generally older than most thin disk stars. As most of the thick disk stars they are on moderately elliptical orbits that typically reach higher distances from the Galactic plane. Type III nebulae should also present underabundances of heavy elements, compared to thin disk stars. Here we assume that the thick disk have a scale height of 1450~pc (Gilmore \& Reid 1983, Reddy et al. 2006).

Finally, type IV PNe are halo objects. Only a few observed objects belong to this class, identified by the low heavy element abundances and the large deviations from disk kinematics (high velocity with respect to the Sun). These nebulae are remnants of very low mass halo stars (a rather uniform old and metal-poor population). They do not show excess in helium, but they appear to have a slight deficiency of this element relative to the other PNe.

There is still a fifth group named type V, which consists of the nebulae found in the Galactic center (Maciel 1989). According to Stasi\'nska et al. (1992), this group is composed by 700 known PNe. Bulge objects have a distinct evolutionary history than the disk and halo and a variety of chemical composition and stellar masses might be related to this component of the Galaxy. For this reason, type V nebulae cannot be defined in terms of the same criteria used to define the types I to IV. Ignoring bulge PNe as a distinct population, however, leads to larger uncertainty in the classification of the other types, as will be discussed separately in \S~\ref{sec:dificulties}. Here we used a method similar to that used by Pottasch \& Acker (1989), Stasi\'nska et al. (1991) and Zhang (1995) to select bulge objects: i) angular radius smaller than 10 arcsec; ii) Galactic longitude ($l$) and latitude ($b$) within $10\degr$ of the Galactic center (which places the nebulae 1.32~kpc above or below the Galactic plane, if $R_\odot = 7.6$~kpc, see Maciel 1993), and iii) flux at 5~GHz less than 100 mJy (Pottasch 1990). Because the criteria used to select type V are not common to those adopted for the types I to IV, this group cannot be used in our statistical analysis (\S~\ref{sec:posclass}).

For types I to IV, the limits between each group are defined by four quantities: the He/H and $\log({\rm N/O})$ abundance ratios, the absolute values of the height above the Galactic disk, $\vert z \vert$, and of the peculiar velocity, $\vert \Delta V \vert$, of each object, following that order of importance.

The height above the Galactic disk, $z$ is given by 

\begin{equation}
z = \dhel\, \sin b \,, \label{z}
\end{equation}

\noindent where \dhel\ is the heliocentric distance of the object. The peculiar velocity is basically the difference between the measured radial velocity relative to the Local Standard of Rest, $V_{\rm LSR}$, and the radial velocity obtained from the Galactic rotation curve (assuming the object has a circular orbit). The planetary nebula radial peculiar velocity is calculated from the equation (Maciel \& Dutra 1992)

\begin{equation}
\Delta V = V_{\rm LSR} - R_\odot \left[ {\Theta(\rgal) \over \rgal} - {\Theta_\odot \over R_\odot} \right] \sin l \cos b \,, \label{Dv}
\end{equation}

\noindent where, \rgal\ is the distance to the Galactic center projected in the Galactic plane,

\begin{equation}
\rgal = \left[ R_\odot^2 + (\dhel\, \cos b)^2 - 2\, R_\odot\, \dhel\, \cos b\, \cos l \right]^{1/2} \,, \label{rgal}
\end{equation}

\noindent $\Theta(\rgal)$ is the rotation velocity at the nebula galactocentric position calculated for a rotation curve. We adopted the rotation curve given by Maciel \& Dutra (1992) for type I + II PNe.

\begin{equation}
\Theta(\rgal) = 314.1356-20.5234\, \rgal + 0.9855\, R_{\rm Gal}^2 \,, \label{theta}
\end{equation}

\noindent For the rotation velocity in the Sun position we adopted $\Theta_\odot = 185$~\kms\, (Rohlfs et al. 1986).

The criteria for classification of the four types proposed by Peimbert (1978), after some reassessments (Peimbert \& Torres-Peimbert 1983; Fa\'undez-Abans \& Maciel 1987a; Maciel 1989; Maciel \& K\"oppen 1994), are shown in Table~\ref{tab:criteria}. The last three columns give, respectively, mean estimates for the progenitor mass and the age intervals corresponding to each type, and the expected Galactic population to which the planetary nebula belongs. For completeness, the type V PNe were also listed in the table.

It should be noted that the masses and ages given in Table~\ref{tab:criteria} are based on the original Peimbert criteria, and do not take into account some more recent attempts to estimate these parameters on the basis of individual methods, as in Maciel et al. (2003, 2005, 2006). Mass intervals in Table~\ref{tab:criteria} should be used as a guide only, since large uncertainties are associated with estimates of progenitor and central star masses. Besides, recent results from Marigo (2007) suggest that hot bottom burning (supposed to be responsible for an overabundance of He and N in the stellar outer layers) may be weakened or even prevented by the third dredge-up process during the early stages of the thermally-pulsating AGB phase in stars within $\approx 3.0$--4.0~\msun\, and metallicities $Z \ga 0.001$, if effects of variable molecular opacities are taken into account. This could reduce the mass range of type I PNe to masses higher than $\approx 3.0$--4.0~\msun.

The resulting classification, following the criteria displayed in Table~\ref{tab:criteria} is shown in the last column of the Table~\ref{tab:data}. In this table, 34\% of the whole sample are type II objects (24\% IIa and 10\% IIb), 28\% are type V and 9\% are type I. Only 3 objects were classified as type III and 7 objects were classified as type IV (each group representing 1\% of the whole sample). Approximately 4\% of the sample (without He/H abundance) were not classified. The rest of the sample (approximately 23\%) consists of objects that could not be fitted in one class, most being between types II, III and IV.

In spite of having those criteria, the classification process is not straightforward since most nebulae do not satisfy all the necessary criteria to fit undoubtedly into a given class. Most likely, they can be classified into two or three types or not even being classified at all. One example is NGC~6807. For this planetary nebula ${\rm He/H} = 0.110$, $\log({\rm N/O}) = -0.60$, $\vert z \vert = 0.66$~kpc, and $\vert \Delta V \vert = 162$~\kms. Its position indicates that this object can be a type IIb nebula. The high peculiar velocity, however, indicates that this object may be a type III or IV. Because there is no limit in abundance among these three groups, the planetary nebula cannot be classified in one single class. 

In an attempt to evaluate how accurately each object fits in its group, some flags were added to the resulting classification. We use an asterisk ``$\ast$'' to indicate absence of $\log ({\rm N/O})$ measurement (in this case the object cannot have an accurate classification). Letter ``A'' indicates that one of the conditions between angular radius smaller than 10 arcsec and flux at 5~GHz less than 100 mJy is not obeyed (applies only for type V nebula). Letter ``B'' indicates that the condition for $\vert z\vert$ is not obeyed. Letter ``C'' indicates that the condition for $\vert \Delta V \vert$ is not obeyed. Letter ``D'' indicates absence of $\vert z\vert$ and/or $\vert \Delta V \vert$ (and/or angular radius and/or flux for the type V nebulae). Letter ``E'' indicates that more than one among the conditions ``A'', ``B'', ``C'' and ``D'' are applied (for types I, IIa and V). Finally, a ``F'' indicates a perfect classification: the source has all characteristics of the type. In the example cited in the last paragraph, the resulting classification of NGC~6807 would be IIbC/(III/IV)B.

\subsection{Reasons for a posterior classification} \label{sec:dificulties}

The merit of the Peimbert classification is in the attempt of sampling PNe in different populations (halo, thin disk, thick disk). Naively, the sequence of types I--II--III--IV could be translated into a sequence of decreasing mass intervals for the planetary nebula progenitor stars in the main sequence, in the sense that type I have high mass progenitors, type II intermediate mass, and type III and IV low and very low masses, respectively. In reality, although distinct mass intervals are usually provided for each of these types, these are only mean estimates, and well defined boundaries between them cannot be established from the four criteria, He/H, N/O, $\vert z \vert$ and $\vert \Delta V \vert$ only. Peimbert himself, in Peimbert \& Carigi (1998), affirmed that it is not possible to have a sharp mass boundary between types II and III based on dynamical arguments only or a sharp mass boundary based on the observed abundances, either. Three causes are provided: i) observational errors in the abundances determinations, ii) displacements of the older PNe from the original Galactocentric distance at which the progenitor star originated, and iii) a real scatter in the chemical abundances of the interstellar medium at a given Galactocentric distance. 

Besides those reasons, additional factors may contribute to difficulties in the classification of PNe into types I to IV: 

\begin{enumerate}

\item {\it Properties that are common to distinct stellar populations}: although the mean properties of the dominant stellar populations of the different components of the Galaxy are reasonably well defined, some characteristics may be common to them (Wyse \& Gilmore 2005). This problem is reflected in the He/H and N/O abundance ratios for types IIb, III and IV PNe. All these types are low metallicity objects, so they are likely to have low He and O abundances. At first a low N abundance would also be expected both because: i) they come from an old stellar population, probably formed from a nitrogen poor interstellar medium; and ii) they are remnants of low mass stars so the nebula should not be self-contaminated by He and N from the stellar nucleosynthesis. Based on this kind of reasoning, we should find progressively lower He, O and N abundances as we go from type IIb to type IV PNe. The interpretation in terms of the N/O ratio, however, is not as simple. 
From the limit of $\log({\rm N/O})$ for type IIb in table \ref{tab:criteria}, one may expect to find $\log({\rm N/O}) < -0.60$ for both thick and halo PNe, but observations show that, at least in regard to the halo population, this is not true. Howard et al. (1997) measured the chemical composition of nine halo PNe, finding that most objects show enhanced $\log (\rm{N/O})$ abundance ratio\footnote{Values range from $-$0.91 to +0.17; only 3 out of the 9 PNe have $\log({\rm N/O}) < -0.60$. All nine objects exhibit subsolar O/H.}. The spread in $\log (\rm{N/O})$ is much larger than can be accounted by uncertainties alone. Unfortunately, the small number of type IV PNe with reliable abundances and distances somewhat hinders their application in statistical studies. From compiled data up to the end of 2001, Stasi\'nska (2004) mentions a total of only 20 known halo PNe (Wyse \& Gilmore 2005; Costa \& Maciel 2006). The large scatter in the N/O abundance ratio in metal-poor stars is also suggested by high precision abundance data for Galactic halo stars (Spite et al. 2005; Israelian et al. 2004). In these papers, the scatter in the N/O abundance ratio is also much larger than their quoted error bars. A model able to predict high N/O abundance ratios at low metallicities was presented in Chiappini et al. (2005), where the explanation for the high nitrogen abundance was based on an increase of the rotational velocity in very metal-poor stars (Meynet \& Maeder 2002). More recent models by Chiappini et al. (2006a, b), considering N yields for stars above 20~\msun\, at metallicity $Z = 10^{-8}$ (Hirschi 2007), suggest that at $\log({\rm O/H})+12 < 7$ (or $[{\rm Fe/H}] < -3$), the main contribution for the larger amounts of N would come from fast rotating massive stars. After $\log({\rm O/H})+12 \sim 7$, where AGB contribution would start to be effective, intermediate mass stars of high rotational velocities could also produce large amounts of nitrogen. In this framework, the scatter in N/O abundance ratios would reflect the distribution of stellar rotational velocities as function of metallicity.
In spite of the uncertainties associated to the observed data, this picture suggests that the interstellar medium from which the halo PNe progenitors have been formed had a significant scatter in N/O. Thus, even considering that the contribution of the halo progenitor star for the N enrichment should not be significant (once these stars are of very low mass) the use of PNe N/O abundance ratio as an accurate tracer of stellar populations is unclear. 
Besides, a real scatter in abundance, another complication arises if the position of the PNe was affected by the motion of the progenitor star away from its original galactocentric birth radius. This possibility is particularly suitable for the older and high velocity PNe. As a consequence, it is possible that among PNe considered to be in the halo, some actually probably belong to an old disk population; also, samples of PNe considered as members of the Galactic disk may actually contain PNe of the halo population that are found physically in the same region as the disk. We may also find PNe of the disk or halo population that are found physically in the same region as the Galactic bulge.

\item {\it Uncertainties in the classification parameters and the absence of a homogeneous data sample}: we verified that typical differences in abundance measurements resulted mostly in changes in the classification between types IIa to I and IIb to IIa. Nevertheless, the number of objects for which this happens is really small. A major source of uncertainty in the classification comes from the adopted distance. The use of a different distance scale does not have a significant effect over the classification of type II and I nebulae, for which the classification criteria are more dependent on chemical abundances. However, some uncertainties may arise in the classification of types IIb, III and IV, for which classification has a larger dependence on distance-related parameters.

\item {\it Objects located in the direction of the Galactic center}: the classification of PNe seen towards lines of sight that are in the direction of the Galactic center has a larger degree of uncertainty than usual. This is due to the broad interval of possible values for the observed radial velocity. Because a variety of radial velocities are allowed to be observed in the direction of the center of our Galaxy, no limit can be established over the peculiar velocity, $\Delta V$, and it cannot be used as a constraint anymore. As a consequence, instead of four parameters (namely He/H, $\log ({\rm N/O})$, $\vert z \vert$, $\vert \Delta V \vert$), only three criteria can be used to classify such objects, increasing uncertainties in the classification. This problem is significantly minimized when we consider PNe from the Galactic center as a distinct population (type V or bulge PNe), and exclude this groups from the sample previous to the classification of the PNe sample into the types I, IIa, IIb, IIII and IV.

\end{enumerate}

\section{Posterior classification of planetary nebulae} \label{sec:posclass}

When one of the classificatory parameters is not available to a given PN, it is usually not possible to associate this object to a single Peimbert type. A more rigorous procedure, based on the observational data available to the PNe, uses the posterior probabilities for an object to be a member of a given type that is defined from the distribution of parameters of a well-defined sample.

We consider an ideal sample for which all PNe are satisfactorily distributed into the Peimbert types, each nebulae having classificatory parameters that satisfy all the conditions imposed by the classificatory criteria. The distribution of classificatory parameters (He/H, $\log \mathrm{N/O}$, $\vert z \vert$ and $\vert \Delta V \vert$) of the population of objects in each group may be estimated by a multivariate parameter distribution of a sample of such group. The probability of each planetary nebula in particular to be member of a given group is calculated by using the multivariate distribution of the group. This method differs from the original classificatory scheme, because it provides a quantitative result of the representativity of the object within its group. Also, through the use of marginal distributions it is possible to extend the Peimbert classification even to those objects that do not have all the necessary classificatory parameters.

In addition to type V PNe, we have also excluded from the analysis those PNe having ${\rm He/H} < 0.060$ and those having simultaneously very high He/H (generally ${\rm He/H} > 0.180$) and very low N/O abundance ratios. In the first case, measurements of ${\rm He/H} < 0.060$ are likely to be errors since these values are lower than the primordial helium abundance (assuming $0.2384$ for the pre-Galactic helium abundance by mass, from Peimbert et al. 2002, which should correspond to an abundance by number of about 0.060). PNe that are observed to have very low He/H abundances should be of low-excitation, having some amount of neutral helium present in the nebula. Because neutral helium cannot be measured, the total helium abundance is obtained from the sum of the He$^+$/H$^+$ and He$^{++}$/H$^+$ ionic abundance only. In those cases, the measured He/H abundance cannot be derived accurately, being underestimated in most cases. For the high He/H abundance ratio, we expect to see a trend of increasing N/O as He/H increases. This trend is commonly explained by dredge-up episodes, which occur in the red giant branch (RGB) and in the AGB phases: in particular, for the most massive stars experiencing hot-bottom burning during the thermal pulses on the AGB phase, the abundance of $^{14}$N should be significantly affected, due to the conversion of C into N in the envelope. The hot-bottom burning yields higher N/O and lower C/O ratios. (Marigo 2001; Perinotto et al. 2004). Finally, it should be noted that, although He2-152 (${\rm He/H} = 0.269$  , $\log ({\rm N/O}) = +0.60$) and M1-11 (${\rm He/H} = 0.094$  , $\log ({\rm N/O}) = +0.77$) do not fit in the group of PNe with simultaneously very high He/H and very low N/O abundance ratios, those PNe are outliers in the trend of increasing N/O versus He/H and for this reason they were also excluded.

Let us consider $\vec x$ a four-dimensional vector composed of the four classificatory parameters of the $K$ Peimbert types: $\vec x = (\mathrm{He/H}, \log \mathrm{N/O}, \vert z \vert, \vert \Delta V \vert)$. The probability of finding a nebula having the set of observables $\vec x$ is:

\begin{equation}
p(\vec x|{\cal I}) = \sum^K_{k=1} p(k|{\cal I}) p(\vec x \vert k,{\cal I}) \,, \label{f(x)}
\end{equation}

\noindent where $p(\vec x \vert k,{\cal I})$ is the likelihood that one can observe $\vec x$ given that the nebula belongs to the group $k$ and $p(k|{\cal I})$ is the prior probability that a given nebula belongs to the group $k$, when no other information is provided. The symbol ${\cal I}$ emphasizes that these probabilities depend on the state of the available background information. 
The prior probability obeys the normalization condition:

\begin{equation}
\sum^K_{k=1} p(k|{\cal I}) = 1  \label{sumpk}
\end{equation}

\noindent and $p(k|{\cal I}) > 0$, $\forall k$. We assume that $p(\vec x \vert k,{\cal I})$ belong to the same parametric family of distributions for all $k$, that is, the likelihood functions have the same profile and are distinguished in each group by a small number of parameters. We assume that $p(\vec x \vert k,{\cal I})$ is a four-dimensional Gaussian, represented by

\begin{equation}
p(\vec x|k,{\cal I}) = (2\pi)^{-2} \left\vert\Sigma_k\right\vert^{-1/2} \exp\left\{ -{1\over 2} 
\left(\vec x - \vec\mu_k\right)^T \Sigma_k^{-1} \left(\vec x - \vec\mu_k\right)\right\} \,, \label{f(x|k)}
\end{equation}
\noindent where $\vec\mu_k = (\mu^k_\mathrm{He/H}, \mu^k_{\log \mathrm{N/O}}, \mu^k_{|z|}, \mu^k_{|\Delta V|})$ is the vector of means, $\Sigma_k$ is the covariance matrix corresponding to group $k$, and $\left\vert\Sigma_k\right\vert$ is its determinant.

The marginal likelihood of a variable $x_q$ may be found by integrating the $p(\vec x|k,{\cal I})$ over all other variables:

\begin{eqnarray}
 p(x_q|{\cal I}) & = & \sum^K_{k=1} \int^{+\infty}_{-\infty} \cdots \int^{+\infty}_{-\infty} p(\vec x|k,{\cal I})
\prod_{i,\, i\ne q}\mathrm{d}x_i \nonumber \\
 \phantom{p(x_q|{\cal I})} & = & \sum^K_{k=1} p(k|{\cal I}) p(x_q|k,{\cal I}) \,\hfil , \label{f(xq)}
\end{eqnarray}
where $i$ stands for the four classificatory parameters used in the Peimbert classification. From Eq. \ref{f(xq)} we see that the marginal likelihood of $x_q$, of a population mixture, is composed by the mixing of the marginal likelihoods of $x_q$ in each of the $K$ groups considered. Also, the marginal likelihood of any observable set $\vec y$ composed by a sub-group of variables of $\vec x$ is given by the integration of $p(\vec x \vert k,{\cal I})$ over the other variables which do not belong to $\vec y$.

Our goal is to estimate the probability that a nebula belongs to a given Peimbert type after considering all its available observational data. According to Bayes Theorem, we can calculate these posterior probabilities as 

\begin{equation}
p(k|\vec y,{\cal I}) = {p(k|{\cal I}) p(\vec y|k,{\cal I}) \over \sum^K_{j=1} p(j|{\cal I}) p(\vec y|j,{\cal I})}\,, \label{p(k|x)}
\end{equation}
where $\vec y$ is the set of known classificatory parameters for the nebula and $p(\vec y|k,{\cal I})$ is the marginal 
likelihood of observing the set $\vec y$ in a nebula belonging to group $k$. Note that 
the posterior probabilities satisfy the equation $\sum^K_{k=1} p(k|\vec y,{\cal I})=1$, $\forall \vec y$. As new observational data are added to the sample, the classification of the nebula can be revised to take into account this new information. In this case the posterior probability found above becomes the prior probability in the right side of Eq. \ref{p(k|x)}, and new posterior probabilities can be calculated.

Here, we assume $K = 5$, representing the five groups of Peimbert I, IIa, IIb, III and IV. At first only PNe that are precisely classified in each of those groups are selected, in order to estimate the likelihood $p(\vec x \vert k,{\cal I})$ of the corresponding group. As a consequence, only nebulae that have all the four classificatory parameters are used.
In the following discussion, we will leave the ${\cal I}$ dependence implicit in order to simplify the notation.

Only seven objects from our sample were unambiguously classified as type IV planetary nebula. In order to improve the statistical analysis of type IV PNe, we have added to this group four other objects that were classified as IVC/IIbB. A similar problem occurs for type III PNe, which has only three unambiguously classified objects. We added to this group 15 objects that were classified as IIIC/IIbB or IIbC/IIIB. This was done for the sole purpose of increasing the sample in order to not completely hinder the statistical analysis for these less populated groups. However, by mixing PNe having uncertain classification with well-classified ones, the estimated multivariate likelihood for types III and IV PNe are surely less well-defined than those for the other types.

%=====================================================================
\begin{table*}
\begin{minipage}[t][]{\textwidth}
%\begin{minipage}[t][180mm]{\textwidth}
\caption{PNe used to define the multivariate likelihood for each Peimbert type}\label{tab:input}
%\centering
\begin{tabular}{ccccccccc}
\hline\hline
\noalign{\smallskip}
\multicolumn{9}{c}{Type I} \\
\noalign{\smallskip}
\hline
\noalign{\smallskip}
BV5-1 & He2-15 & He2-143 & He2-153 & K3-70 & M1-8 & M1-51 & M1-75 & NGC 2899 \\
NGC 6302 & NGC 6445 & NGC 6537 & NGC 6781 & NGC 6803 & NGC 7293 &  &  &  \\
\hline
\noalign{\smallskip}
\multicolumn{9}{c}{Type IIa} \\
\noalign{\smallskip}
\hline
\noalign{\smallskip}
He2-29 & He2-37 & He2-51 & He2-86 & He2-119 & He2-140 & He2-141 & IC 418 & IC 2149 \\
IC 2448 & IC 2501 & IC 2621 & K3-68 & M1-4 & M1-7 & M1-13 & M1-57 & M1-58 \\
M1-60 & M1-61 & M1-74 & M1-79 & M1-80 & M2-2 & M2-4 & M2-46 & M2-55 \\
M3-1 & M3-4 & M3-5 & M3-29 & MyCn18 & Mz1 & Mz2 & NGC 650 & NGC 2022 \\
NGC 2346 & NGC 2371-72 & NGC 2438 & NGC 2440 & NGC 2452 & NGC 2792 & NGC 2867 & NGC 3132 & NGC 3195 \\
NGC 3699 & NGC 3918 & NGC 5189 & NGC 5315 & NGC 5882 & NGC 6572 & NGC 6720 & NGC 6741 & NGC 6818 \\
NGC 6881 & NGC 6884 & NGC 7026 & NGC 7027 & NGC 7354 & Th2-A    &    &    &   \\
\hline
\noalign{\smallskip}
\multicolumn{9}{c}{Type IIb} \\
\noalign{\smallskip}
\hline
\noalign{\smallskip}
H3-75 & Hb12 & He2-115 & He2-47 & He2-55 & Hu1-1 & Hu2-1 & IC 2003 & IC 2165 \\
IC 351 & IC 4191 & IC 4406 & IC 4634 & IC 4776 & IC 5117 & J900 & K2-1 & M1-14 \\
M1-17 & M1-5 & M1-50 & M1-6 & M3-54 & M3-6 & NGC 40 & NGC 3211  & NGC 3242 \\
NGC 4361 & NGC 6210 & NGC 6309 & NGC 6578 & NGC 6790 & NGC 6826 & NGC 6879 & NGC 6891 & NGC 6905 \\
NGC 7662 & Pe1-1 & Tc1 & & & & & & \\
\hline
\noalign{\smallskip}
\multicolumn{9}{c}{Type III} \\
\noalign{\smallskip}
\hline
\noalign{\smallskip}
Fg1 & He2-5 & He2-21 & He2-67 & He2-99 & He2-149 & He2-157 & He2-158 & IC 4846 \\
J320 & KFL 19 & M1-9 & M2-9 & M2-53 & M3-33 & NGC 1535 & NGC 6326 & PC14 \\
\hline
\noalign{\smallskip}
\multicolumn{9}{c}{Type IV} \\
\noalign{\smallskip}
\hline
\noalign{\smallskip}
BoBn1 & H4-1 & He2-118 & IC 3568 & IC 4593 & K648 & PC12 & DdDm1 & Me2-1 \\
Sn1 & Vy1-2 &   &  &  &  &  &  &   \\
\noalign{\smallskip}
\hline
\end{tabular}
\vfill
\end{minipage}
\end{table*}
%=====================================================================

%=====================================================================
\begin{table*}
\caption{Statistic parameters of $p(\vec x \vert k)$}
\label{tab:statistic}
\centering
\begin{tabular}{lrrrrr}
\hline\hline
\noalign{\smallskip}
    & Type I & Type IIa & Type IIb & Type III & Type IV \\
\noalign{\smallskip}
\hline
\noalign{\smallskip}
$\mu(\mathrm{He/H})$                       &  0.1503  & 0.1135 & 0.1035 & 0.1004 & 0.0983 \\
$\mu(\log \mathrm{N/O})$                   & $-$0.0160 & $-$0.3973 & $-$0.8110 & $-$0.9161 & $-$0.7873 \\
$\mu(|z|)$ (kpc)                         &  0.0812 & 0.2251 & 0.3761 & 0.6859 & 5.4504 \\
$\mu(|\Delta V|)$ (\kms)                 & 25.1667 & 28.1200 & 23.3615 & 69.6389 & 115.5273 \\
\noalign{\smallskip}
\hline
\noalign{\smallskip}
$c(\mathrm{He/H}, \mathrm{He/H}) $         & 0.0003 & 0.0003 & 0.0002 & 0.0003 & 0.0001 \\
$c(\mathrm{He/H}, \log \mathrm{N/O})$      & 0.0015 & $-$0.0005 & 0.0001 & $-$0.0001 & 0.0023 \\
$c(\mathrm{He/H}, |z|) $                     & $-$0.0002 & $-$0.0005 & $-$0.0000 & $-$0.0026 & 0.0277 \\
$c(\mathrm{He/H}, |\Delta V|) $              & 0.0168 & $-$0.0395 & $-$0.0146 & 0.0642 & 0.0143 \\
$c(\log \mathrm{N/O}, \log \mathrm{N/O})$  &  0.0444 & 0.0373 & 0.0300 & 0.0275 & 0.2231 \\
$c(\log \mathrm{N/O}, |z|) $                 & $-$0.0062 & $-$0.0071 & $-$0.0013 & $-$0.0331 & 1.7941 \\
$c(\log \mathrm{N/O}, |\Delta V|) $          & $-$0.7432 & 0.0729 & $-$0.2483 & 1.2787 & 14.5046 \\
$c(|z|, |z|) $                               & 0.0034 & 0.0445 & 0.0663 & 0.2045 & 25.8489 \\
$c(|z|, |\Delta V|) $                        & $-$0.0391 & 0.1964 & 0.2333 & $-$5.9943 & 218.0222 \\
$c(|\Delta V|, |\Delta V|) $                 & 203.3552 & 282.4165 & 269.3030 & 495.8237 & 8091.3482 \\
\noalign{\smallskip}
\hline
\noalign{\smallskip}
$p(k)$                                      & 0.105 & 0.419 & 0.273 & 0.126 & 0.077 \\
\noalign{\smallskip}
\hline
\end{tabular}
\end{table*}
%=====================================================================

\begin{figure*}
%\sidecaption
%\includegraphics[width=12cm]{fig1.eps}
\resizebox{\hsize}{!}{\includegraphics{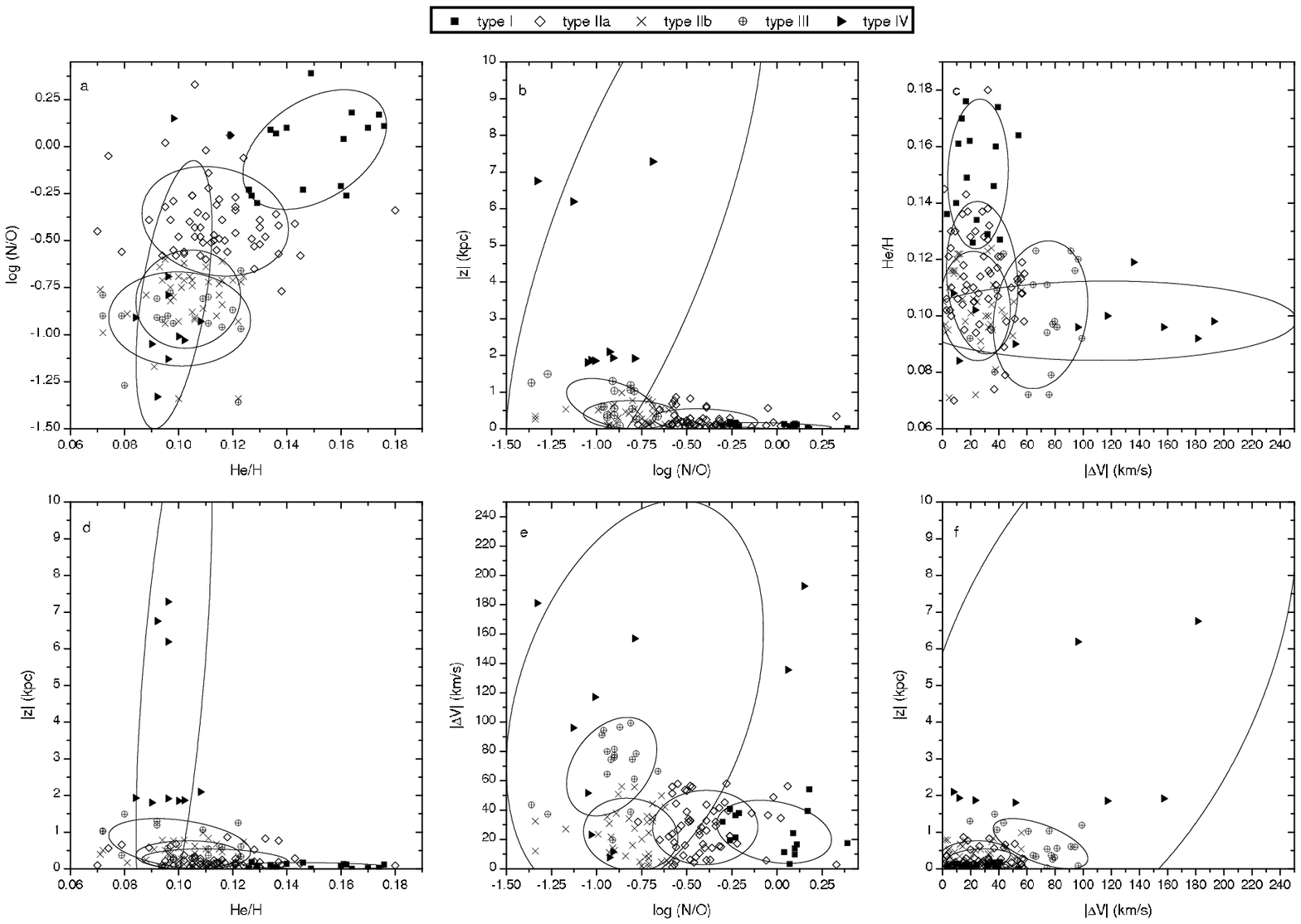}}
\caption{Projections of the multivariate distribution of classificatory parameters in the Peimbert classification scheme. Different symbols are used for PNe unambiguously classified in the five main Peimbert group, as shown in the caption. An ellipse marks the 1-sigma confidence level of each group bivariate distribution in the four-dimensional space (He/H, $\log \mathrm{N/O}$, $|z|$, $|\Delta V|$). While some groups are reasonably well separated in some of the plots, particularly He/H $\times\log \mathrm{N/O}$ and $\log \mathrm{N/O} \times |\Delta V|$, they overlap considerably in other plots, on account of the ambiguity in the Peimbert classification and of real overlaps of some of the parameters that characterize the different nebular populations.}
\label{figgrupos}
\end{figure*}

In Table~\ref{tab:input}, we list the PNe used in the calculation of $p(\vec x \vert k)$ to each of the Peimbert types. In Table~\ref{tab:statistic}, we list the statistical parameters of $p(\vec x \vert k)$. The mean of the variable $x_i$ is represented by $\mu(x_i)$, and the notation $c(x_i, x_j)$ indicates the covariance between the variables $x_i$ and $x_j$. Accordingly, $c(x_i,x_i)$ is the variance of $x_i$. Note that $c(x_i,x_j) = c(x_j,x_i)$. Figure~\ref{figgrupos} shows two-dimensional projections of the four-dimensional likelihoods $p(\vec x \vert k)$ overimposed on the data for the PNe listed in Table~\ref{tab:input}. The ellipses mark the 1$\sigma$ confidence level for the bivariate distribution of the parameters plotted, for each of the $K$ groups. In this figure, the role of each classificatory parameter can be considered. He/H and $\log \mathrm{N/O}$ are discriminant parameters for the earlier Peimbert types (I to II), but cannot separate between IIb to IV PNe types. To separate these types, we need to consider $|z|$ and $|\Delta V|$, which, on the other hand, are not good discriminant parameters for the earlier types. We have verified that the Peimbert groups do not naturally arise as optimally decomposed groups of the four-dimensional variable mixing when a proper statistical method for the decomposition of population mixing is applied to the data\footnote{We have used the {\tt EMMIX} software provided by McLachlan et al. (1999)}. That is, the Peimbert classification is rather {\em ad hoc}, in the sense that it is based on the apriori expectation that objects within given parameter ranges are members of given stellar populations of the Galaxy.

The shape of the ellipses in Figure \ref{figgrupos} reveals some interesting properties of the Peimbert types. For instance, there is a trend of finding larger $\log \mathrm{N/O}$ and $|\Delta V|$ with increasing $|z|$ among type IV PNe. Because these PNe are thought to be members of the Galactic halo, we can conclude that this trend should be found among halo stars as well. The increase of $|\Delta V|$ with increasing $|z|$ is indeed verified among halo stars. It indicates that stars with increasing height above the Galactic plane have a tendency to progressively lag behind the Local Standard of Rest. On the other hand the $\log \mathrm{N/O}$--$|z|$ trend found for these nebulae reinforces that metal-poor stellar populations could have a high N/O abundance ratio, as discussed in section \ref{sec:dificulties} above. Other trends revealed by the ellipses are somewhat unexpected, as the decrease in $|\Delta V|$ for increasing $|z|$ among type III PNe. It is presently unclear whether this trend is real or artificially imposed into the classification of type III PNe by the fuzzy limit in $|z|$ for the separation among type III to type IV PNe in the original Peimbert Classification. This trend deserves a further consideration after more data becomes available.

%=============================================================================
%=============================================================================

\begin{figure*}
\resizebox{\hsize}{!}{\includegraphics{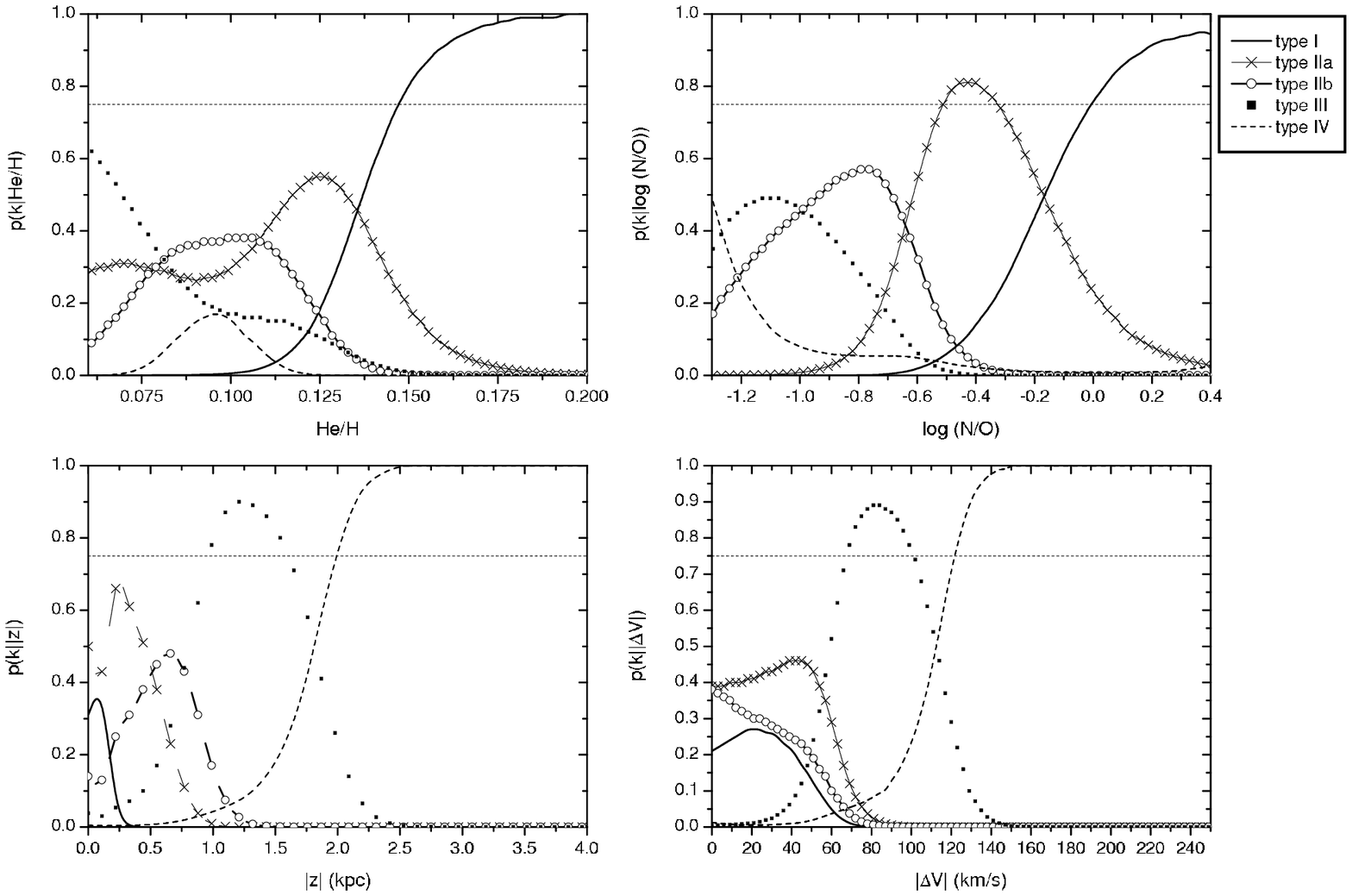}}
\caption{Posterior probability, $p(k \vert x_q)$, or the fraction of PNe which are likely to be members part of a group $k$, when only one single information for the nebula is known. The horizontal line indicates $p(k \vert x_q$) = 0.75. According to the figure, $\log \mathrm{N/O}$ and $|z|$ are the most important parameters in the Peimbert classification.}
\label{mixingport}
\end{figure*}

The marginal likelihoods tell us about how each variable affects the classification of the nebulae. In Fig.~\ref{mixingport} we show how the posterior probabilities $p(k \vert x_q)$ depends on each variable. In those plots, the horizontal line correspond to $p(k \vert x_q) = 0.75$, that we arbitrarily assumed as the limit for a safe classification of a nebulae in a given group. In a classification system that has few ambiguities, we expect to find almost each curve rising above $p(k \vert x_q) = 0.75$ in a given variable range. Contrary to this expectation, Fig.~\ref{mixingport} shows that the several $p(k \vert x_q)$ curves overlap each other considerably for some variables. As a consequence, only a small number of curves do reach $p(k \vert x_q) > 0.75$. This indicates that Peimbert classification is rather ambiguous if only one classificatory parameter is known. However, the same problem affects other population classification schemes in Astronomy, like the Galactic stellar population classification (see, for instance, Nemec \& Nemec 1993).

Seven intervals can be defined from these curves, within which the PNe can be classified with reasonable accuracy from a single parameter. Those are summarized in Table~\ref{tab:intervals}. We conclude that $\log \mathrm{N/O}$ is an important determinant parameter in the classification of the types I and IIa, with two intervals of unambiguous classification in Table~\ref{tab:intervals}. $|z|$ and $|\Delta V|$ are also discriminatory parameters between Type III and IV PNe, but not for those younger nebulae associated with the thin disk. He/H is not an unimportant parameter, but the overlap between the several $p(k \vert {\rm He/H})$ curves hinders a secure nebulae classification unless He/H $> 0.147$. 

%=============================================================================
%=============================================================================

More narrow intervals can be defined by considering that only two or three classificatory parameters are known. We do not show the corresponding plots in this case, since no new information can be taken from them that is not already present in the Figs. \ref{figgrupos} and \ref{mixingport}. If more than one classificatory parameter is known, it is best to use Eq. \ref{p(k|x)} to fully classify the nebula.

%=====================================================================
\begin{table}
\caption{Best parameter intervals for safe classification of PNe}
\label{tab:intervals}
%%\centering
\begin{tabular}{cl}
\hline\hline
\noalign{\smallskip}
Interval & Class \\
\noalign{\smallskip}
\hline
\noalign{\smallskip}
$\mathrm{He/H} > 0.147$                & type I PNe \\
$\log \mathrm{N/O} > -0.005$           & type I PNe \\
$-0.513 < \log \mathrm{N/O} < -0.324$  & type IIa PNe \\
$0.95 < |z| < 1.60\, \mathrm{kpc}$      & type III PNe \\
$|z| > 1.99\, \mathrm{kpc}$            & type IV PNe \\
$66 < |\Delta V| < 101\,~\kms$         & type III PNe \\
$|\Delta V| > 121\,~\kms$              & type IV PNe \\
\noalign{\smallskip}
\hline
\end{tabular}
\end{table}
%=====================================================================

The accuracy of the likelihoods $p(\vec x \vert k)$ used in the posterior classifications depends on the accuracy of the prior classification that defined the `fiducial' PNe samples. We had to incorporate some ambiguously classified Type III and IV PNe to these samples in order to have a meaningfully large sample. On account of this, some ambiguity is likely to have been incorporated into $p(\vec x \vert k)$. A glimpse of Fig.~\ref{figgrupos} shows that the present `fiducial' Type III PNe sample comprises PNe with lower He/H abundance than the `fiducial' Type IV PNe sample, which are supposed to be older than Type III PNe and, thus, more He poor. This problem can be eliminated once the number of prior classified nebulae increases as more data becomes available.\footnote{Recently, Pereira \& Miranda (2007) identified PNG~232.0$+$05.7 as a halo object, increasing the small group of known halo planetary nebulae (PNG~232.0$+$05.7 present a very low N/O abundance ratio, with $\log ({\rm N/O}) = -0.89$). Analogously, PNG~034.5-11.7 was identified as a type III planetary nebula. Unfortunately none of these new identified PNe have He/H measurements.}

In Table~\ref{tab:probab} (available also in electronic form) we list the posterior probability $p(k \vert \vec y)$ for the membership of each planetary to the Peimbert groups, where $\vec y$ is the vector composed by the classificatory parameters of the nebula that are known. For comparison purpose, we also list the pre-classification, as in Table~\ref{tab:data}. {\it While the posterior classification agrees reasonably well with the prior classification for those PNe used in the `fiducial' samples, the novelty of this result is that we can now give a quantitative estimate of how likely each ambiguously classified nebula belongs to a given Peimbert group}. This result is particularly important in the case of nebulae for which a few classificatory parameters are known. The resulting posterior classification listed in the eighth column corresponds to the group with higher posterior probability. Compared to the percentages provided in the \S~\ref{sec:peimbert} for the numbers of PNe in each group, we have now that, excluding the objects that were pre-classified as bulge PNe (28\% of the whole sample), 42\% of the whole sample of 476 PNe are type II PNe (31\% IIa and 11\% IIb), 11\% are type III, 9\% are type I, and 6\% are type IV. There are still a number of PNe with less than 50\% of chance to pertain to a certain group, including 3 objects that still cannot be classified, having equal probabilities to belong to two different types. Those PNe represent 4\% of the sample.

In order to avoid indiscriminate use of the results presented in Table~\ref{tab:probab}, we wish to present some critical comments. First of all, we find that a number of 30 PNe are now classified as halo type IV. Those include objects classified as halo PNe in the literature (Howard et al. 1997) plus objects with ambiguous pre-classification among two or more Peimbert types. We have noticed, however, that this group contains 11 PNe previously classified as type IIa. Although the scale heights for most of these objects are less than 1 kpc, their peculiar velocities are high. The inclusion of these objects in the type IV group may be partially explained by the large scatter in N/O observed in halo objects. Most of the pre-classified IIa PNe have low He/H ($< 0.125$) but $\log({\rm N/O})$ $> -0.60$.

It also should be noted that in Table~\ref{tab:probab} we list posterior classification for type V bulge PNe. This was done in order to investigate the main characteristics of this group, given the meaning of the sequence of types I to IV as being a sequence of intervals of progenitor mass (see discussion ahead). We do not recommend the use of our resulting posterior classification for those PNe pre-classified as type V for application in studies dealing with Galactic evolution or related purposes. In those case, bulge objects should be used as a separate group.

{\it Planetary nebula bulge population}: It is interesting to investigate the characteristics of the PNe in the bulge population using the posterior classification of PNe that have a type V prior classification. Until now, we have discarded from the statistical analysis objects belonging to this class. This procedure has assured a better characterization of the types I to IV. Here we can study this class separately. We calculated the posterior classification of all PNe with prior classification VA, VD and VF in Table~\ref{tab:data}. Those are listed in Table~\ref{tab:probab}. For this group of objects we have 6 type I, 41 type IIa, 19 type IIb, 35 type III and 32 type IV PNe. We conclude that type V PNe are, indeed, a mixed group comprising a variety of chemical composition and stellar masses. They are likely to be misclassified in all other Peimbert groups if the proper classificatory parameters for type V PNe are not taken into account. The frequency of misclassified type V PNe as Type III or IV PNe, however, compared to the prior probabilities $p(k)$ is larger by a factor of at least 2. This suggests that most type V PNe are more similar to type III and IV PNe, in agreement with what we would expect from Galactic population studies, according to which the bulge population is closer to the thick disk and halo population than to the thin disk population (Feltzing \& Gilmore 2000; Ferreras et al. 2003).

\section{Discussion and Application} \label{sec:discuss}

In the Figs~\ref{figgrupos} and \ref{mixingport} we note significant overlap among the distributions of the classificatory parameters, which best define the different types I to IV. As commented in the previous sections, the lack of sharp boundaries between these intervals is one of the main causes for uncertainties in the classification of PNe into these types. The overlap in the $p(\vec x \vert k)$ curves should be reduced with a better characterization of the sample, which can be reached with the improvement of the observational data, taking into account new and more precise measurements. A larger number of PNe pre-classified into the types III and IV is also of major importance, since those groups are poorly populated in our analysis. We do not expect, however, that the overlaps in Figs. \ref{figgrupos} and \ref{mixingport} will be completely removed. Some overlaps are probably real, since different nebula populations have common properties. Some discussion was already presented in \S~\ref{sec:dificulties} concerning the large scatter observed in abundances from thin disk to halo, and possible effects of the motion of the progenitor stars in the position of the PNe. Thus, overlaps should exist not only in the curves of the He/H and $\log {\rm N/O}$ abundance ratios, but also, although in a smaller range of coincidence, in the curves for $|\Delta V|$ and $|z|$, particularly for the older PNe.

This discussion reinforces our idea that statistical methods can greatly contribute to a better characterization of groups of PNe. It is also clear that these groups cannot be well defined using a limited number of criteria, and the consideration of further quantities that can be related to stellar evolution, kinematics or spatial distribution can probably help in the characterization of stellar populations. Following this line of thought, Phillips (2005) has evaluated trends in several planetary nebula properties for the Peimbert types I, IIa, IIb, and III, finding marked differences in properties among the four Peimbert classes, and confirming a likely differentiation in progenitor and ionized shell masses. Among the quantities analyzed were the: nebular radii, expansion velocity, density, shell ionized masses, morphology, radial velocity, Galactic latitude, 5~GHz brightness temperature, Zanstra temperature (likely to be related to central star effective temperature), central star luminosity, He$^{++}$/He$^+$, dust-to-gas mass ratios, and molecular-to-ionized mass ratios. In the following subsections we present additional comments on the relation between morphological classes and PNe Peimbert types. We also briefly comment on isotopic ratios as indicators of the mass of planetary nebula progenitor star.

\subsection{Planetary nebula morphological class as a stellar population indicator} \label{sec:morphology}

Several studies use planetary nebula morphology as indicator of stellar populations. It was first noticed by Peimbert \& Torres-Peimbert (1983) that a large fraction of type I PNe (He and N rich) show bipolar structure (see also Calvet \& Peimbert 1983). Later on, the association of progenitors of bipolar PNe and the higher-end mass range of the AGB stars was confirmed in a series of papers dealing with the correlation of the planetary nebula morphology and properties of the nebula and its central star (Amnuel 1995; Corradi et al. 1997; Corradi \& Schwarz 1995; G\'orny et al. 1997; Stanghellini et al. 1993, 2002). Morphological classes are usually defined on the basis of the H$\alpha$, [\oiii]$\,\lambda\,5007$ or [\nii]$\,\lambda\,6584$ images. Attempts to establish a morphological classification have been done by many authors (Greig 1972; Zuckerman \& Aller 1986; Balick 1987; Chu et al. 1987; Schwarz et al. 1992; Manchado et al. 1996). Some variation in the definition of groups and subgroups exist, depending on the level of detail which is taken into account, by considering or not additional structures and morphological features (inner structures, ansae, rings, multiple shell structures, and other). Also, different interpretation of a given image, may result in differences in the classification assigned by different authors.

\begin{figure}
\resizebox{\hsize}{!}{\includegraphics{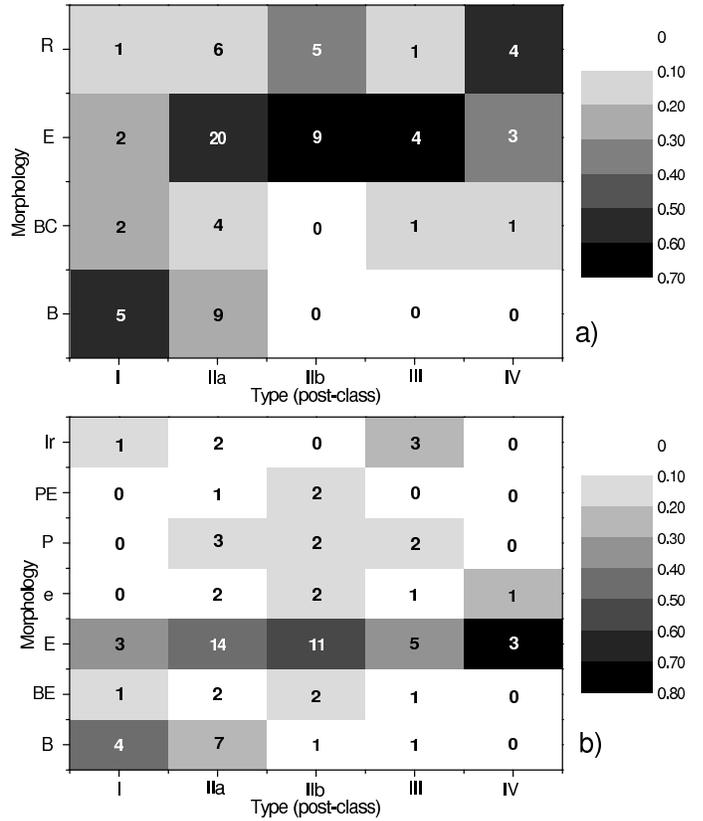}}
\caption{Distribution of morphological classes and PNe types (according to our posterior classification) for: a) 77 objects in common with Stanghellini et al. (2002) and b) 77 objects in common with G\'orny et al. (1997) and Stasi\'nska et al. (1997). The scales in the captions are normalized numbers: in the upper plot (a), groups were normalized by 10 type I, 39 IIa, 14 IIb, 6 III and 8 IV PNe; while in the lower plot (b) we have 9 type I, 31 IIa, 20 IIb, 13 III and 4 IV PNe.}
\label{fig:morphology}
\end{figure}

Fig.~\ref{fig:morphology}a shows a comparison between the Peimbert types and the morphological classes in Stanghellini et al. (2002) for PNe in common with their study. Fig.~\ref{fig:morphology}b is analogous to Fig.~\ref{fig:morphology}a, but now showing PNe in common with G\'orny et al. (1997) and Stasi\'nska et al. (1997). Peimbert types are given according to our posterior classification. The number of objects in each interval are plotted. For a proper comparative analysis, the groups are normalized by the total number of PNe in each of the Peimbert types. The gray color scale refers to the normalized numbers, where darker colors indicate higher numbers of objects. In Fig.~\ref{fig:morphology}a we have 10 type I PNe, 39 IIa, 14 IIb, 6 III and 8 IV. In Fig.~\ref{fig:morphology}b we have 9 type I, 31 IIa, 20 IIb, 13 III and 4 IV, summing a total number of 77 objects in each figure.

%=============================================================================
%=============================================================================

In G\'orny et al. (1997) and Stasi\'nska et al. (1997), PNe were classified as point-symmetric (P), bipolar (B), elliptical (E) and irregular (Ir). Further subdivisions were given to these groups, with the subdivision of the point-symmetric into PE and P, according to whether the knots are embedded or not in an overall structure. Analogous classification was applied to the bipolar, resulting in the BE and B sub-groups. The elliptical were subdivided into {\it e}, elliptical disk without structure; E, clear elliptical morphology; and EH, almost circular with very regular shape and bright rim (we have none of this last morphological class).

Stanghellini et al. (2002) used a more simplified morphological scheme, taking into account three main morphological classes, namely round (R), elliptical (E) and bipolar (B), each morphological classes being associated to different stellar populations. Some details were included by considering the bipolar core group (BC). According to this classification, round PNe should be the remnants of the lower mass progenitors, elliptical planetary nebulae represent low to intermediate mass progenitors, and bipolar PNe are the result of high stellar mass progenitor evolution (Stanghellini et al. 2002, 2006; Manchado 2004). Hereafter, we will refer to this classification as Simplified Morphological Classification.

From Fig.~\ref{fig:morphology}a, most of the type I correspond to bipolar objects, most of the types IIa, IIb and III, are elliptical and most of the type IV are round PNe, which roughly confirms the expected correlation between the morphological classes and the stellar population. A similar conclusion can be drawn from the application of a contingency table test for the null hypothesis that the Peimbert and the Simplified Morphological Classification are independent, using the data from Fig.~\ref{fig:morphology}a. We have verified that this null hypothesis can be rejected with a 0.05 level of significance ($P$-value = 0.03), i.e., the PNe classes in the Simplified Morphological Classification are not entirely uncorrelated with those classes from the Peimbert Classification. However, there is no straight correspondence between these morphological classes and the Peimbert types. For example, for 17 nebulae in common with the round morphological group, 6 are IIa, which is not consistent with the supposition that round PNe were predominantly in the low-mass domain. 

In Fig.~\ref{fig:morphology}b, for the G\'orny et al. (1997) plus Stasi\'nska et al. (1997) sample, the interpretation of correspondence between the classifications is not clear, probably due to the many subdivisions in the morphological classes. Some speculation, however, can be done about this plot. Point-symmetric PNe are supposed to be present in a variety of main-body morphologies, which implies that in the Simplified Morphological Classification, they could be found in the round, bipolar or elliptical classes (Stanghellini et al. 2002). Stanghellini et al. (2002) call attention to a similarity in the definition of bipolar core (BC) and the one used by G\'orny et al. (1997) for embedded bipolar (BE), which, also, would be classified as R or E in the Simplified Morphological Classification. Roughly, we find that Type I objects are almost equally distributed in the E and B classification. For all the other Peimbert types, the E morphology dominates. Since G\'orny et al. (1997) do not distinguish between round and elliptical, it is possible that some of the type IV PNe could be classified as round. Thus, although less obviously, the distributions in Fig.~\ref{fig:morphology}b still can be interpreted as showing a relation of morphological class and PNe mass progenitor, but the variety of morphological structures found inside of each of the types I to IV avoid strong statements. A contingency table test for the independence between the G\'orny et al. morphological classification and the Peimbert Classification reinforces our conclusions: The $P$-value for rejecting the null hypothesis in this test is 0.52, indicating that both classification schemes have high probability of being independent.

A N/O $\times$ He/H plot of elliptical, round and bipolar PNe would show significant overlap between different morphological samples. One example can be found  in Stanghellini et al. (2006, Fig. 1), which shows elliptical and round PNe in the same data range, including some in the typical data range for bipolar PNe. Thus, although we do find indication that the main morphological classes, namely bipolar, elliptical and round are in some way correlated to the Peimbert types, which are supposed to correspond to a sequence of stellar populations of the Galaxy, it is clear that PNe morphological class cannot be used as unique indicator of stellar population.

\subsection{The {\rm $^{12}$C/$^{13}$C} isotopic ratio as a stellar mass indicator} \label{sec:isotopabund}

The abundance content of several chemical elements and isotopic ratios ($^3$He, $^4$He, $^{12}$C/$^{13}$C, $^{14}$N/$^{15}$N, $^{16}$O/$^{17}$O) in PNe results from complex combination of different dredge-ups, mixing, mass loss and nuclear processing episodes that occur along the evolution of their progenitors. In particular, abundances that will be significantly affected should be those of He, C, and N (and possibly O). Those changes will depend mainly on the initial stellar mass and, in a minor degree, on the metallicity. This leads to the possibility of using observations of isotopic abundance ratios in PNe as constraints of stellar mass and age.

Standard models predict that in specific phases of the stellar evolution, the external convective envelope moves downward reaching internal layers where H burning has occurred, and bring up processed material to the stellar surface. As a result, changes in the surface abundance will occur after each dredge-up episode. A significant increase in the surface $^{12}$C/$^{13}$C isotopic ratio is expected after the third dredge up in an AGB, where the H burning shell is expected to be enriched with $^{12}$C due to a previous He shell flash. The third dredge up results in high $^{12}$C/$^{13}$C ratios, larger than 20 (Boothroyd \& Sackmann 1999; Renzini \& Voli 1981). As observed values of the $^{12}$C/$^{13}$C isotopic ratio on the RGB and later evolution (Charbonnel \& Nascimento 1998) suggest that values predicted by standard models are overestimated, additional processes, employing deep extra mixing mechanisms during RGB and/or AGB phase are usually applied in order to explain discrepancies (Charbonnel 1994, 1995; van den Hoek \& Groenewegen 1997; Marigo 2001). Those non-standard processes result in a significant enhancement of the $^{13}$C abundance in the surface layers.

%=============================================================================
\begin{figure}
\resizebox{\hsize}{!}{\includegraphics{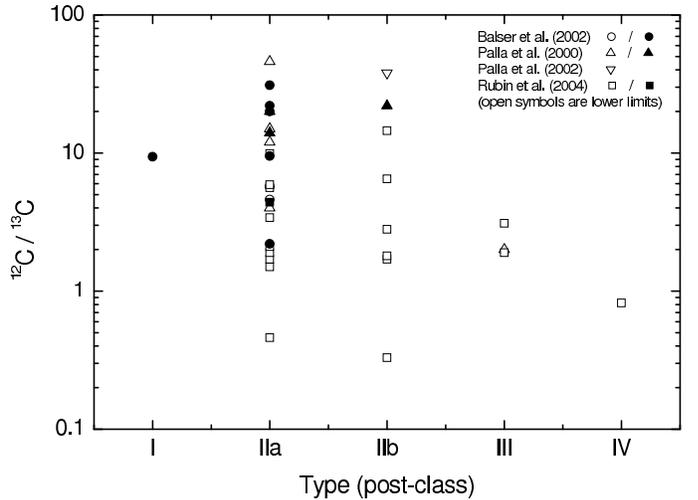}}
\caption{The $^{12}$C/$^{13}$C isotopic ratio in the different Peimbert types, according our post-classification. Regardless uncertainties in observational data and theoretical models, large scatters are possible for the isotopic ratio in each type.}
\label{fig:c12c13}
\end{figure}
%=============================================================================

In low mass stars ($\la 2$~\msun), nonstandard mixing models that incorporate cool bottom processing (during RGB and/or AGB phase) are usually applied to reproduce observational data. In this process the bottom of the convective envelope remains cool while an ad hoc mixing mechanism brings material down to layers hot enough for some nuclear processing. Such models can lead the $^{12}$C/$^{13}$C isotopic ratio to values as low as 4--5 (Wasserburg et al. 1995). For masses larger than $\approx 4$~\msun, stars develop hot bottom burning in the AGB phase, where deep convective envelop with very high base temperature activates the CN cycle, as it penetrates into the H burning shell, processing $^{12}$C into $^{13}$C and $^{14}$N. This process tend to drive $^{12}$C/$^{13}$C toward $\sim 3.3$ (Smith \& Lambert 1990; Frost et al. 1998). 

In Fig.~\ref{fig:c12c13}, we present the posterior classification of PNe with observed $^{12}$C/$^{13}$C isotopic ratios. Data were taken from Balser et al. (2002), supplemented by Palla et al. (2000, 2002) and Rubin et al. (2004). For Palla et al. (2000) and Balser et al. (2002) measurements were made from millimeter transitions of $^{12}{\rm CO}$ and $^{13}{\rm CO}$. These samples comprise a larger number of type I and mostly type IIa PNe. Later types, comprehending PNe originated from older stars, are found in larger (although still small) number in the studies from Palla et al. (2002) and Rubin et al. (2004), which used the \ciii] multiplet, containing the $^{12}$C lines at 1906.7 and 1908.7~\AA\, and the very weak $^{13}$C line at 1909.6~\AA. For a large number of PNe, in which the $^{13}$C line was not detected, only lower limits of the $^{12}$C/$^{13}$C isotopic abundance ratios were provided. Most of these PNe are elliptical or bipolar (one bipolar core), regardless of the origin of the sample.

The types I to IV should be interpreted as a sequence of decreasing mass for the PNe progenitor star in the main sequence. Those estimates should provide a critical check of the Peimbert classification as a sequence of mass intervals. It is beyond the scope of this paper to analyze these types as a function of stellar mass, since there are large uncertainties associated with those estimates, and no interesting constraint (no new information) would be provided, in addition to the ones already discussed in Balser et al. (2002) and referred works that used mass estimates. Some attempts have already been made, in those studies, to establish a relationship of the $^{12}$C/$^{13}$C isotopic ratio in PNe with the mass of the progenitor star in the main sequence, but large uncertainties in the determination of the stellar mass and inconsistencies in standard stellar evolution theory and observations hinder any strong statements (Balser et al. 2002). The sequence of types in Fig.~\ref{fig:c12c13} should provide an independent qualitative picture of any possible trend of the $^{12}$C/$^{13}$C ratio with the mass of the progenitor star. 

In the figure, the observations of the CO millimeter-wave and \ciii] transition in PNe yield $^{12}$C/$^{13}$C values between $\sim 2.2$ and 31, with progenitor masses from $\sim 1$ to 4~\msun. Lower limit data expand these intervals to $\sim 0.5$ and 45, roughly, for the same mass interval. Low values (lower than 5) are observed for the $^{12}$C/$^{13}$C isotopic ratios in the older types III and IV, in agreement with results provided by models, that use deep mixing cool bottom processing. Because a large number of IIb PNe may have masses lower than 2~\msun, cool bottom burning should also occurs in, at least, part of the group of PNe classified as type IIb. In the intermediate mass interval, comprising types IIa and IIb, a large scatter in the $^{12}$C/$^{13}$C ratio is observed, suggesting that only part of the stars suffer deep mixing. In fact, a problem remains for stars between 2--4~\msun, since cool bottom processing, is only expected to occur in progenitor stars of $\leq 2$~\msun. Thus, low $^{12}$C/$^{13}$C between 2 and 4~\msun\, might indicate further processing on the AGB or that the observed ratios are underestimated. The type I planetary nebula in the figure shows a low $^{12}{\rm C}/^{13}{\rm C}$ ratio, in agreement with models that use hot bottom burning. This extra mixing mechanism, however, should occur only in the more massive ($\ga 4$~\msun) objects in the type I mass range. Because the type I PNe group also contain stars with masses lower than 4~\msun, we should also observe higher values for the $^{12}$C/$^{13}$C ratio within this group.

In summary, we could expect to observe a trend of increasing $^{12}$C/$^{13}$C ratio in the direction of the lower masses type I PNe, with some scatter. A large scatter should be observed within the types IIa and IIb, with a slight trend of decreasing $^{12}$C/$^{13}$C ratios in the direction of the lower mass IIb PNe. Because some scatter in the isotopic ratio should always be observed in the types I to IIb, the only significant constraint the $^{12}$C/$^{13}$C isotopic ratio could provide would be for the types III and IV, which are expected to present low values. These conclusions, however, cannot be taken for granted, since larger uncertainties are still associated with stellar evolution models and observational data. Observations of NGC~3242 (estimated mass of 1.2~\msun) by Palla et al. (2002) resulted in a carbon isotopic ratio $> 38$, in agreement with standard stellar models. This could provide evidences that not all progenitors undergo a phase of deep mixing during the last stages of its evolution. Based on chemical evolution models, more than 90\% of low-mass stars must undergo cool bottom processing (Charbonnel et al. 1998). Thus, NGC~3242 must belong to the small group of standard low-mass stars, consisting of $\sim 10$\% of stars with mass lower than 2~\msun. Additionally, a large fraction of the data presented is lower limits. A large number of observational measurements, in particular from \ciii] transition, in order to cover older PNe, is desirable. Concerning theory, new models that incorporate the effects of axial rotation (Meynet \& Maeder 2002; Meynet et al. 2006) or of variable molecular opacities (opacities consistent with the changing photospheric chemical composition) in the competition between third dredge-up and hot bottom burning during thermally pulsating AGB phase (Marigo 2007; Marigo \& Girardi 2007) can provide progressive advances in the present understanding of stellar evolution in the intermediate mass range.

\section{Summary}\label{sec:concl}

We have re-analyzed the criteria used to characterize the Peimbert types I, IIa, IIb, III and IV, by performing a Bayesian statistical classification of a large sample of PNe previously classified into these groups. Our sample consists of 476 PNe, whose nebular properties have been compiled from selected sources from the literature. A subsample, composed only by PNe satisfactorily distributed into the types I to IV, was used to calculate the joint probability density function of the classificatory parameters (He/H, $\log \mathrm{N/O}$, $\vert z \vert$ and $\vert \Delta V \vert$) in the five groups considered. These distributions were then used to calculate the probability of each planetary nebula to be member of a given Peimbert type. This probability, called posterior probability, provides a quantitative result of the representativity of the object within its group. 

Our posterior classification increases the number of PNe classified into the Peimbert types, extending the classification to PNe that are ambiguously classified in the traditional method. Even PNe for which a few classificatory parameters are known can be classified. 

Uncertainties in our posterior classification are given by the posterior probabilities for an object to be member of each of the types I, IIa, IIb, III and IV. The classification can be improved as new observational data is added to the sample, and the classification of the nebula is revised, to take into account this new information.

Our analysis confirms that: i) The types I and IIa are well defined in terms of abundance, contrary to the other three groups IIb, III and IV. In particular, $\log{\rm N/O}$ is more relevant to define the different types. ii) All the types are relatively well characterized by the height, $\vert z \vert$, which clearly is one of the most important parameters to classify a halo PNe. iii) The peculiar velocity tends to increase as we go from thin disk, thick disk to halo, as expected. iv) $\vert z \vert$ and $\vert \Delta V \vert$ are essential parameters in the classification of the different components of the Galaxy (thin disk, thick disk and halo), although they cannot be used alone to distinguish the three types pertaining to the thin disk (I, IIa and IIb).

We found that one of the main causes for ambiguities existing in the original method of selection of PNe into the types I to IV should be associated with the difficulty in defining sharp mass boundaries for each of these groups. This conclusion can be inferred from the superposition we have found of the curves of distributions of the classificatory parameters (He/H, $\log \mathrm{N/O}$, $\vert z \vert$ and $\vert \Delta V \vert$), which best define the different types I to IV. The overlap in the probability density functions will be reduced as the observational data is improved by taking into account new and more precise measurements (not available for all objects) and including a larger number of PNe pre-classified into the types III and IV, since those groups are poorly populated in our analysis.

The overlaps in the curves of distributions of He/H, $\log \mathrm{N/O}$, $\vert z \vert$ and $\vert \Delta V \vert$, however, should be in part explained by real overlap of some properties of different stellar populations. This suggests that the classification can be improved if a larger number of classificatory parameters is taken into account.

We have checked how morphological PNe classifications compare to the Peimbert Classification. We have found that the Simplified Morphological Classification used in Stanghellini et al. (2006) is not uncorrelated with the Peimbert Classification. There is a trend of having more bipolar type I PNe, elliptical type II and III PNe and round type IV planetary nebula in the Galaxy. This indicates that PNe morphology may be an indicator of the stellar population to which the planetary nebula belongs. However, the more detailed morphological classification by G\'orny et al. (1997) has been found to be independent from the chemokinematical Peimbert Classification, probably on account of the larger variety of morphologies considered.

Finally, we have used the posterior PNe classification developed in this paper to explore whether there is a trend between the $^{12}$C/$^{13}$C ratio and the mass of the progenitor star, which is likely to be correlated with the Peimbert types if these correspond to a sequence of stellar populations of the Galaxy. Our results do not rule out the existence of this trend, although the available data still prevent us from drawing strong conclusions.

\begin{acknowledgements}
We thank the referee for the remarks that helped us to improve this paper. C.Q. would like to thank Cristina Chiappini for valuable comments on the N/O abundance ratio and Roberto Costa for comments on the He/H abundance. This work was partly supported by the Funda\c c\~ao de Amparo \`a Pesquisa do Estado de S\~ao Paulo (FAPESP) and Conselho Nacional de Desenvolvimento Cient\'{\i}fico e Tecnol\'ogico (CNPq/MCT).
\end{acknowledgements}

%\clearpage
%% TABLES:------------------------------------------------------------------
%% \input{tables.tex}

%=====================================================================
% Table 1 available electronically only
\onllongtabL{1}{
\begin{landscape}
\begin{longtable}{llrlrlrlrrlrrrrrlrll}
\caption{\label{tab:data} Parameters and pre-classification (full table and reference lists are available electronically)}\\
\hline\hline
\noalign{\smallskip}
Name & PNG & He/H & Ref. & $\epsilon({\rm O})^{\mathrm{\dagger}}$ & Ref. & $\epsilon({\rm N})^{\mathrm{\dagger}}$ & Ref. &  $\log({\rm N/O})$ & \dhel & Ref. & \rgal & $\vert z \vert$ & $V_{\rm LSR}^{\mathrm{\ddagger}}$ & $\vert \Delta V \vert$ & $\Theta_{\rm diam}$ & Ref. & $S_{\rm 5~GHz}$ & Ref. &  Type \\
     &     &      &      &     &      &            &     &       & (kpc) &  & (kpc) & (kpc) & (km/s)                                                                                                                                                                                                                                                                                                 & (km/s)     & (arcsec) &      & (mJy) &     &       \\
\noalign{\smallskip}
\hline
\noalign{\smallskip}
\endfirsthead
\caption{continued.}\\
\hline\hline
\noalign{\smallskip}
name & PNG & He/H & ref. & O/H$^{\mathrm{\dagger}}$ & ref. & N/H$^{\mathrm{\dagger}}$ & ref. & $\log({\rm N/O})$ & \dhel & ref. & \rgal & $\vert z \vert$ & $V_{\rm LSR}^{\mathrm{\ddagger}}$ & $\vert \Delta V \vert$ & $\Theta_{\rm diam}$ & ref. & $S_{\rm 5~GHz}$ & ref. &  Type \\
     &     &      &      &     &      &            &     &       & (kpc) &  & (kpc) & (kpc) & (km/s)                                                                                                                                                                                                                                                                                                 & (km/s)     & (arcsec) &      & (mJy) &     &       \\
\noalign{\smallskip}
\hline
\noalign{\smallskip}
\endhead
\noalign{\smallskip}
\hline
\endfoot
A4             &    144.3$-$15.5      &  \dots    &                 \dots    &  \dots    &                    \dots    &  \dots    &                 \dots    &  $-$0.17    &  6.1    &    10    &  12.8    &   1.630    &     \dots    &   \dots    &    20.0    &        27    &     1.5    &       9    &                  indef.      \\
A12            &    198.6$-$06.3      &  0.119    &                     3    &   8.93    &                        3    &   8.06    &                     3    &  $-$0.87    &  2.0    &     4    &   9.5    &   0.219    &     \dots    &   \dots    &    37.0    &        18    &    36.0    &       9    &                    IIbD      \\
A18            &    216.0$-$00.2      &  0.152    &                     3    &   7.99    &                        3    &   7.97    &                     3    &  $-$0.02    &  1.6    &     4    &   8.9    &   0.006    &     \dots    &   \dots    &    73.0    &         7    &    17.0    &       9    &                      ID      \\
A20            &    214.9$+$07.8      &  0.125    &                     3    &   8.80    &                        3    &  \dots    &                 \dots    &  \dots      &  2.0    &     4    &   9.3    &   0.271    &     \dots    &   \dots    &    67.0    &         7    &     7.0    &       9    &               (I/IIa)D*      \\
A24            &    217.1$+$14.7      &  \dots    &                 \dots    &  \dots    &                    \dots    &  \dots    &                 \dots    &  $+$0.43    &  0.3    &     7    &   7.8    &   0.076    &    $+$0.9    &    14.0    &   354.8    &        27    &    36.0    &       9    &                  indef.      \\
A35            &    303.6$+$40.0      &  \dots    &                 \dots    &  \dots    &                    \dots    &  \dots    &                 \dots    &  $-$0.10    &  0.2    &    17    &   7.5    &   0.129    &    $-$5.9    &    15.1    &   772.0    &        27    &   255.0    &       9    &                  indef.      \\
A50            &    078.5$+$18.7      &  0.089    &                    21    &  \dots    &                    \dots    &  \dots    &                 \dots    &  $-$0.40    &  2.8    &    10    &   7.5    &   0.898    &  $-$145.2    &   175.2    &    27.0    &        27    &     1.0    &       9    &                    IIaC      \\
A65            &    017.3$-$21.9      &  0.260    &                    22    &   8.18    &                       22    &   7.33    &                    22    &  $-$0.85    &  1.5    &    10    &   6.3    &   0.559    &   $+$21.8    &     1.9    &   104.0    &        27    &     4.0    &       9    &                    IIaF      \\
A70            &    038.1$-$25.4      &  0.180    &                    22    &   7.98    &                       22    &   7.52    &                    22    &  $-$0.46    &  3.5    &    10    &   5.5    &   1.501    &   $-$69.0    &   145.0    &    42.0    &        27    &    12.0    &       9    &                    IIaE      \\
A71            &    084.9$+$04.4      &  \dots    &                 \dots    &  \dots    &                    \dots    &  \dots    &                 \dots    &  $+$0.38    &  0.9    &    10    &   7.6    &   0.069    &     \dots    &   \dots    &   158.0    &        27    &    82.8    &       9    &                  indef.      \\
A77            &    097.5$+$03.1      &  \dots    &                 \dots    &   8.41    &                       21    &   7.02    &                    21    &  $-$1.39    &  1.5    &    10    &   7.9    &   0.081    &  $-$103.4    &   122.5    &    65.8    &        27    &   307.6    &       9    &                  indef.      \\
A82            &    114.0$-$04.6      &  \dots    &                 \dots    &  \dots    &                    \dots    &  \dots    &                 \dots    &  $-$0.28    &  2.0    &    10    &   8.6    &   0.160    &   $-$24.6    &    25.4    &    81.0    &        27    &     5.3    &       9    &                  indef.      \\
.              &                      &           &                          &           &                             &           &                          &             &         &          &          &            &              &            &            &              &            &            &                              \\
.              &                      &           &                          &           &                             &           &                          &             &         &          &          &            &              &            &            &              &            &            &                              \\
.              &                      &           &                          &           &                             &           &                          &             &         &          &          &            &              &            &            &              &            &            &                              \\
\end{longtable}
%\vfill

\begin{list}{}{}
\item[$^{\dagger}$] $\epsilon({\rm X}) = \log ({\rm X/H}) + 12$.
\item[$^{\ddagger}$] Radial velocities from the catalog of Durand et al. (1998). For every object, the heliocentric radial velocities have been converted to the Local Standard of Rest radial velocities (see text).
\end{list}{}{}

\noindent References of abundances.
(3) Costa et al. 2004;
(21) Perinotto 1991;
(22) Perinotto et al. 1994; \dots

\noindent References of distances.
(4) Costa et al. 2004 (see references in);
(7) Harris et al. 1997;
(10) Maciel 1984;
(17) Pottasch 1996; \dots

\noindent References of the angular diameters.
(7) Cahn \& Kaler 1971;
(18) Perek \& Kohoutek 1967;
(27) Zhang 1995 (see references in); \dots

\noindent References of the 5 GHz flux densities.
(9) Zhang 1995 (see references in); \dots

\end{landscape}
}% End onllongtabL
%%=====================================================================

%\clearpage

%%=====================================================================
% Table 6 available electronically only
\onllongtab{6}{
\begin{longtable}{lrrrrrll}
\caption{\label{tab:probab} Posterior probabilities for the planetary nebulae to belong to Peimbert classes}\\
\hline\hline
\noalign{\smallskip}
Name & $p({\rm type~I} \vert \vec y)$ & $p({\rm type~IIa} \vert \vec y)$ & $p({\rm type~IIb} \vert \vec y)$ & $p({\rm type~III} \vert \vec y)$ & $p({\rm type~IV} \vert \vec y)$ & Pre-class & Post-class \\
\noalign{\smallskip}
\hline
\noalign{\smallskip}
\endfirsthead
\caption{continued.}\\
\hline\hline
\noalign{\smallskip}
Name & $p({\rm type~I} \vert \vec y)$ & $p({\rm type~IIa} \vert \vec y)$ & $p({\rm type~IIb} \vert \vec y)$ & $p({\rm type~III} \vert \vec y)$ & $p({\rm type~IV} \vert \vec y)$ & Pre-class & Post-class \\
\noalign{\smallskip}
\hline
\noalign{\smallskip}
\endhead
\hline
\endfoot
\hline %new
\endlastfoot %new
A4               &  0.00  &  0.00  &  0.00  &  0.00  &  1.00  &                        indef.  &        IV \cr
A12              &  0.00  &  0.11  &  0.67  &  0.22  &  0.00  &                          IIbD  &       IIb \cr
A18              &  0.98  &  0.02  &  0.00  &  0.00  &  0.00  &                            ID  &         I \cr
A20              &  0.00  &  0.78  &  0.16  &  0.06  &  0.00  &                     (I/IIa)D*  &       IIa \cr
A24              &  1.00  &  0.00  &  0.00  &  0.00  &  0.00  &                        indef.  &         I \cr
A35              &  0.75  &  0.25  &  0.00  &  0.00  &  0.00  &                        indef.  &         I \cr
A50              &  0.00  &  0.00  &  0.00  &  0.00  &  1.00  &                          IIaC  &        IV \cr
A65              &  0.00  &  1.00  &  0.00  &  0.00  &  0.00  &                          IIaF  &       IIa \cr
\end{longtable}
}% End onllongtab
%%=====================================================================

\end{document}